\documentclass[onecolumn]{natureprintstyle}

\usepackage[whole]{bxcjkjatype}

\makeatletter
\providecommand{\@LN}[2]{}
\makeatother

\usepackage{amsmath}
\usepackage{amssymb}
\usepackage{amsfonts}
\usepackage{graphicx}
\usepackage{siunitx}
\usepackage{booktabs}
\usepackage[super,sort,comma]{natbib}
\usepackage{multibib}
\usepackage[version=4]{mhchem}
\usepackage{hyperref}
\usepackage[symbol]{footmisc}
\usepackage{threeparttable}

\newcommand\micron{\mbox{$\mu$m}}
\newcommand\ion[2]{\text{#1\,\textsc{\lowercase{#2}}}}

\newcites{mt}{ }
\newcites{si}{ }
\title{The Great Dimming of Betelgeuse seen by the Himawari-8 meteorological satellite} 

\author{Daisuke Taniguchi${}^{1,\dag }$, Kazuya Yamazaki${}^{2}$ \& Shinsuke Uno${}^{3}$}

\begin{document}

\maketitle

\renewcommand{\figureautorefname}{Fig.}
\renewcommand{\tableautorefname}{Fig.}

\renewcommand{\figurename}{Fig.}
\renewcommand{\tablename}{Table}

\begin{affiliations}
\item Department of Astronomy, School of Science, The University of Tokyo, Tokyo, Japan
\item Department of Earth and Planetary Science, School of Science, The University of Tokyo, Tokyo, Japan
\item Institute of Astronomy, School of Science, The University of Tokyo, Tokyo, Japan
\end{affiliations}
${}^{\dag }$\href{mailto:d.taniguchi.astro@gmail.com}{d.taniguchi.astro@gmail.com}

\begin{abstract}
Betelgeuse, one of the most studied red supergiant stars\cite{Montarges2021,Ohnaka2011}, dimmed in the optical by $\mathbf{{\sim }1.2}$\,mag between late 2019 and early 2020, reaching an historical minimum\cite{Guinan2019,Guinan2020,Sigismondi2020} called ``the Great Dimming.'' 
Thanks to enormous observational effort to date, two hypotheses remain that can explain the Dimming\cite{Montarges2021}: a decrease in the effective temperature\cite{Dharmawardena2020,Harper2020b} and an enhancement of the extinction caused by newly produced circumstellar dust\cite{Levesque2020,Dupree2020}. 
However, the lack of multi-wavelength monitoring observations, especially in the mid infrared where emission from circumstellar dust can be detected, has prevented us from closely examining these hypotheses. 
Here we present $\mathbf{4.5}$-year, $\mathbf{16}$-band photometry of Betelgeuse between 2017--2021 in the $\mathbf{0.45\text{--}13.5\,\micron }$ wavelength range making use of images taken by the Himawari-8\cite{Bessho2016} geostationary meteorological satellite. 
By examining the optical and near-infrared light curves, we show that both a decreased effective temperature and increased dust extinction may have contributed by almost the same amount to the Great Dimming. 
Moreover, using the mid-infrared light curves, we find that the enhanced circumstellar extinction actually contributed to the Dimming. 
Thus, the Dimming event of Betelgeuse provides us an opportunity to examine the mechanism responsible for the mass loss of red supergiants, which affects the fate of massive stars as supernovae\cite{Beasor2021}. 
\end{abstract}

\begin{figure}[b!]
\includegraphics[width=0.595\textwidth ]{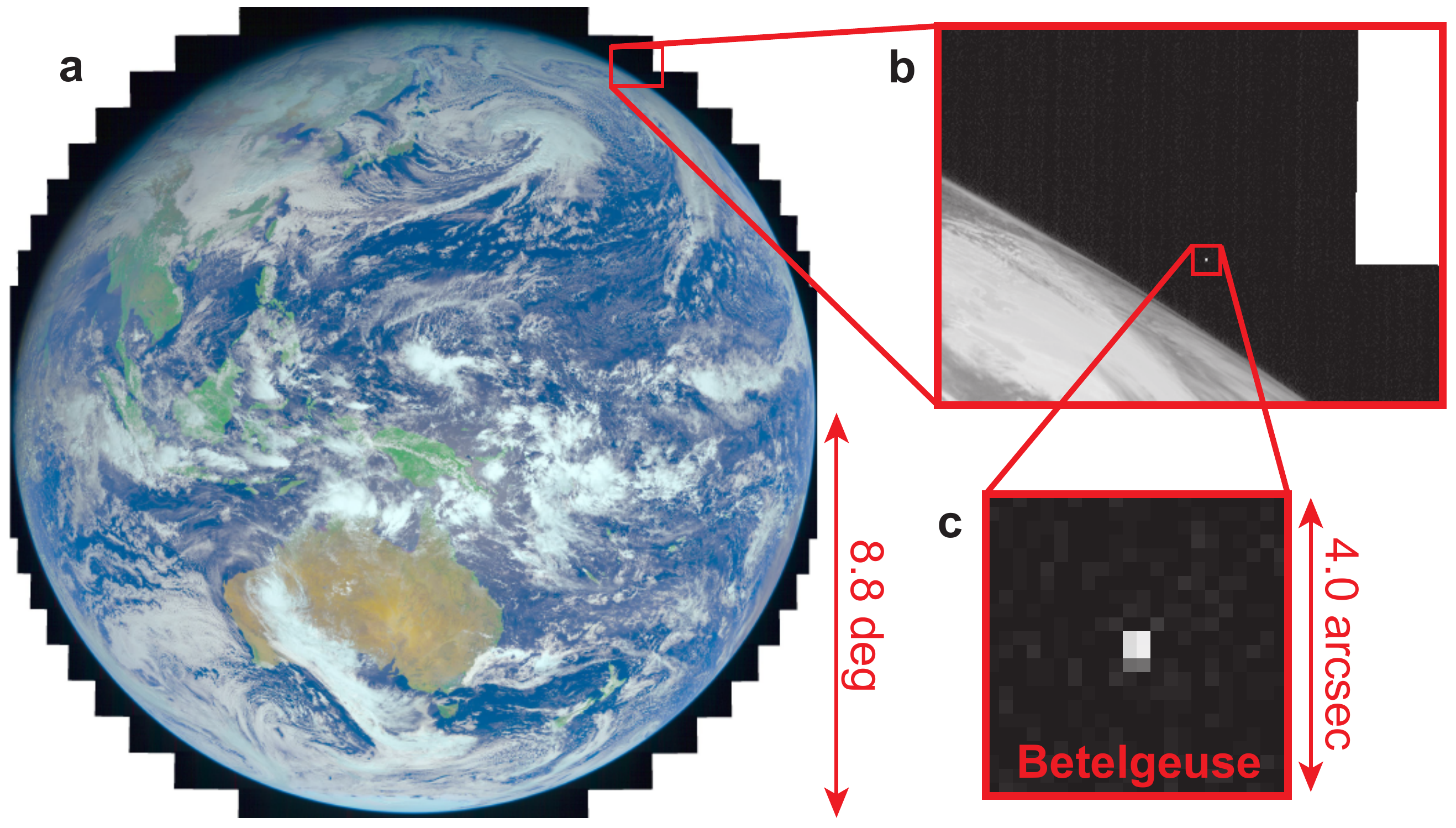}
\includegraphics[width=0.405\textwidth ]{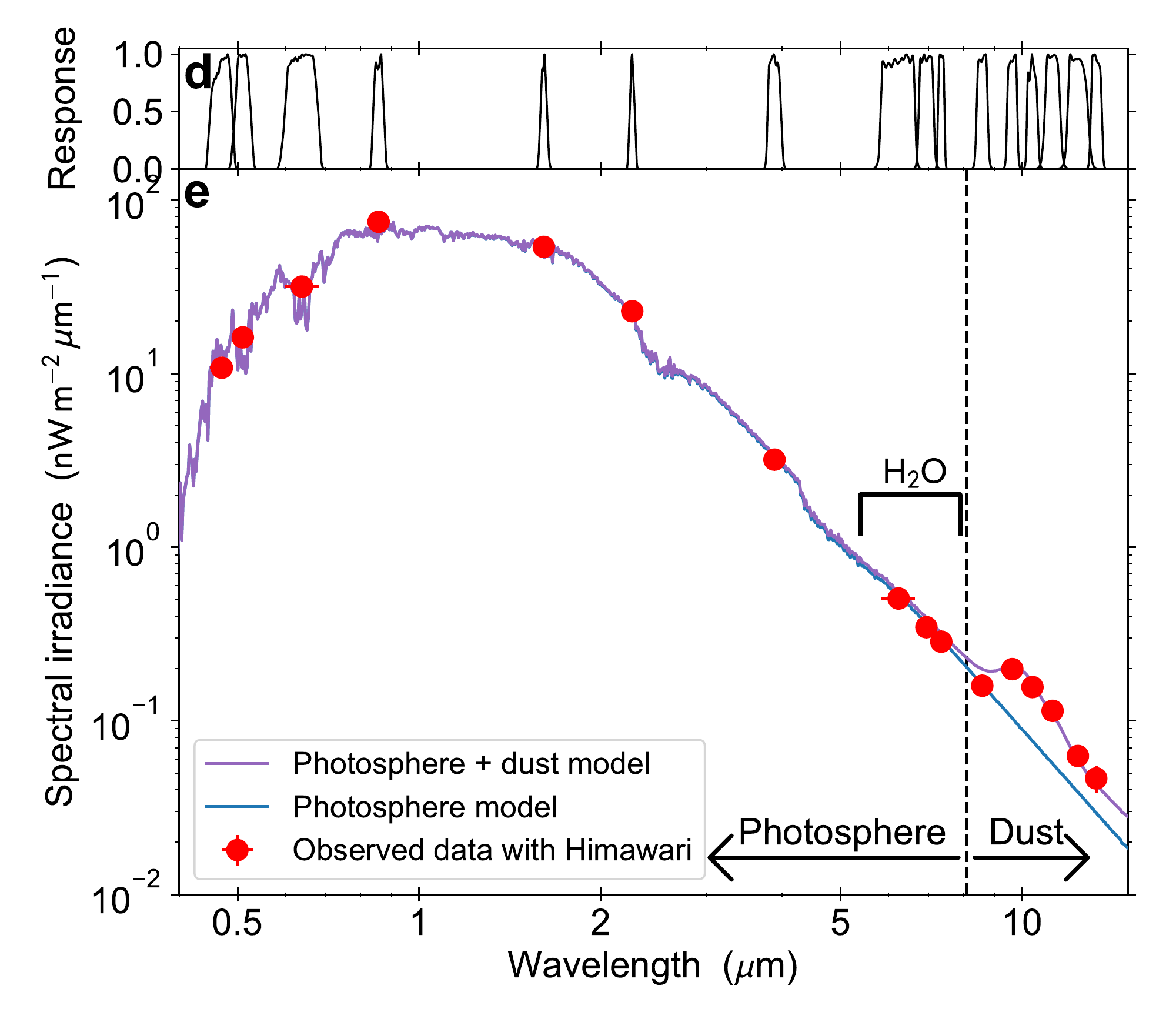}
\caption{\textbf{Betelgeuse observed with the Himawari-8 meteorological satellite near the edge of the Earth's disk. } \textbf{a}, False-color image taken by the Himawari-8 satellite on 19 January 2020. Betelgeuse is located at the top right corner of the image (see zoomed-in images of Band~7 in \textbf{b} and \textbf{c}). \textbf{d}, The response functions of Himawari-8; see the basic properties of the Himawari-8 data in Extended Data \autoref{table:AHI}. \textbf{e}, The observed SED after averaging all observations with a model spectrum fitted to it. }
\label{fig:fulldisk/SEDalldate}
\end{figure}

Himawari-8 is a Japanese geostationary meteorological satellite orbiting $35,786\,\mathrm{km}$ above the equator at $140.7^{\circ}$ east\cite{Bessho2016}.
Since 7 July 2015, Himawari-8 has taken images of the entire disk of the Earth once every ten minutes using its optical and infrared imager, the Advanced Himawari Imager~(AHI; Extended Data \autoref{table:AHI}). 
The Himawari-8 satellite also observes the region of outer space around the edge of the Earth's disk during every scan; this motivated us to develop a brand-new concept: using meteorological satellites as ``space telescopes'' for astronomy~(\autoref{fig:fulldisk/SEDalldate}). 

One of the unique aspects of this concept to use the Himawari-8 for astronomy is that it enables us to obtain high-cadence time series of mid-infrared images, which is hard to acquire with the usual astronomical instruments. 
Small ground-based telescopes can acquire time-series measurements\cite{Gehrz2020}, but the observations only cover wavelengths within the atmospheric window, and they are interrupted by the Sun for several months in most cases\cite{Dupree2020b,Nickel2021}. 
In contrast, astronomical satellites and aircraft-carrying telescopes can observe stars in most wavelength ranges\cite{Tsuji2000,Harper2020a,Harper2021}, but the observations require higher costs than ground-based telescopes. 
Survey telescopes orbiting the Earth for non-astronomical purposes---like the Himawari-8 meteorological satellite---have the potential to overcome these problems. 
In this work, we demonstrate that the mystery of the Great Dimming of Betelgeuse can be solved with the aid of the time-series observations from the Himawari-8, especially that in the mid infrared, with which the amount of dust around Betelgeuse can be determined. 

\twocolumn

Examining the images taken by the Himawari-8 satellite, we found that several bright stars, including Betelgeuse, sometimes appear in the images~(Extended Data \autoref{table:targets}). 
We therefore measured the amount of light coming from Betelgeuse, typically once per $1.72\,\mathrm{days}$, between January 2017 and June 2021, and we made a $4.5$-year catalog of light curves in $16$ bands covering $0.45\text{--}13.5\,\micron $~(Extended Data \autoref{fig:allbands_flux:Betelgeuse}\textbf{a}). 
Comparing the light curves, i.e., the time series of spectral energy distributions~(SEDs), with a model SED for Betelgeuse, we determined the time variation of the main parameters of Betelgeuse, namely the radius $R$, effective temperature $T_{\mathrm{eff}}$, and extinction $A(V)$ in the optical. 
In addition to these three parameters, which can be determined from optical or near-infrared observation\cite{Levesque2020,Taniguchi2021}, we also succeeded in determining the dust optical depth $\tau _{10}$ at $10\,\micron $ using our mid-infrared light curves. 
With this $\tau _{10}$, we will examine the dust-production history around Betelgeuse.

\subsection{Time variations of the photospheric parameters}\label{ssec:TimeSeries}

\begin{figure}[t!]
\includegraphics[width=\columnwidth ]{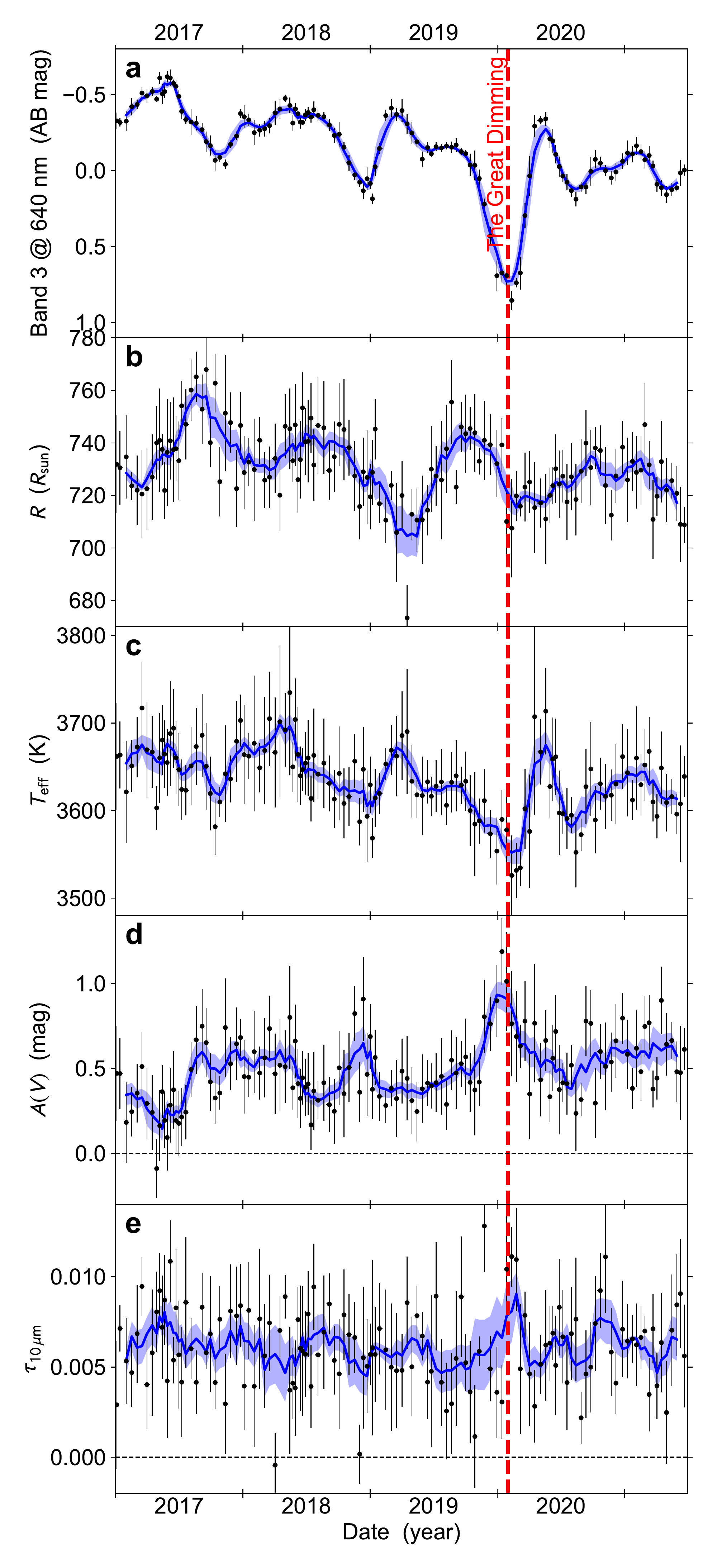}
\caption{\textbf{Time variation of the stellar parameters of Betelgeuse. } The black dots with error bars represent the measured values and their standard errors, and the blue lines show smoothed versions by averaging five points to guide the eye. \textbf{a}, The light curve of Betelgeuse at Band~3, centered at $640\,\mathrm{nm}$, in AB magnitudes. \textbf{b}--\textbf{e}, The results of the SED fitting. \textbf{b}, Radius, $R$. \textbf{c}, Effective temperature, $T_{\mathrm{eff}}$. \textbf{d}, Dust extinction in the \textit{V} band, $A(V)$. \textbf{e}, Dust optical depth at $10\,\micron $, $\tau _{10}$. The time of the Great Dimming is indicated by the red vertical line. }
\label{fig:Timeseries}
\end{figure}

\autoref{fig:Timeseries} shows the main results of this paper, namely, the time variations of the stellar parameters, $R$, $T_{\mathrm{eff}}$, $A(V)$, and $\tau _{10}$ together with the AB magnitudes for Band~3 at $640\,\mathrm{nm}$, which is located at around the central part of the Johnson's \textit{R}-band. 
From the time variations of $R$, $T_{\mathrm{eff}}$, and $A(V)$, we found that the Great Dimming of Betelgeuse, the drop of ${\sim }1.2\,\mathrm{mag}$ in the \textit{V} magnitude\cite{Guinan2020}, is likely to be caused by a cooling of ${\sim }140\,\mathrm{K}$ in $T_{\mathrm{eff}}$ which results in a dimming of the \textit{V} magnitude by ${\sim }0.6\,\mathrm{mag}$, and by an increase of ${\sim }0.6\,\mathrm{mag}$ in $A(V)$ which, by definition, darken the \textit{V} magnitude by ${\sim }0.6\,\mathrm{mag}$. 
This result supports a previously suggested scenario in which the Dimming was caused by a combination of a decrease in $T_{\mathrm{eff}}$ and an increase in $A(V)$\cite{Levesque2020,Montarges2021}, although several shortcomings in our model SED should be kept in mind; some of the shortcomings are examined in the Methods section and Extended Data \autoref{fig:MARCS4par}. 

However, the cause of the enhanced extinction $A(V)$ during the Great Dimming is still under debate. 
Several works have proposed that occultation by a newly formed circumstellar dust cloud might be a promising cause\cite{Levesque2020,Montarges2021}. 
In contrast to this dust-formation scenario, several other works have argued that the dust formation is unnecessary to explain the Dimming. 
For example, three-dimensional~(3D) effects\cite{Harper2020b} or molecular opacity\cite{Kravchenko2021,Davies2021}, rather than dust extinction, can also explain the Dimming. 

In order to examine whether or not the enhanced extinction $A(V)$ during the Dimming is due to the circumstellar dust, we focus on the time variation of $\tau _{10}$. 
Since $\tau _{10}$ was determined using the mid-infrared light curves of Bands~12--14 at around $10\,\micron $, where O-rich dust emissions dominate the flux\cite{Verhoelst2006,Kervella2011}, it reflects the amount of circumstellar dust that emits the infrared light. 
Therefore, using our infrared $\tau _{10}$ measurements, in contrast to optical diagnostics, we can directly trace the amount of circumstellar dust. 
Our measurements in the bottom panel of \autoref{fig:Timeseries} shows an enhancement of $\tau _{10}$ during the Great Dimming, though there is a slight time delay between the peaks of $A(V)$ and $\tau _{10}$. 
This enhancement of $\tau _{10}$ indicates that a clump of gas did in fact produce dust that shaded the photosphere of Betelgeuse and contributed to the Great Dimming. 

Two potential dust reservoirs have been found so far for red supergiants. 
The first is a diffuse and spread-out dust layer located a few to several tens of a stellar radius from the photosphere\cite{Verhoelst2006,Kervella2011}. 
In addition, recent very-high-spatial-resolution instruments have found circumstellar dust clouds around red supergiants just a few tenths of stellar radii above the photosphere\cite{Haubois2019,Asaki2020}. 
The distribution and amount of dust in the spread-out layer remain mostly unchanged with time\cite{Kervella2011,Montarges2020}, and thus it is unlikely that dust condensation in this layer is related to the large time variation of $A(V)$ and $\tau _{10}$ during the Dimming that we found. 
In contrast, the latter, lukewarm region\cite{Ohnaka2013} coexists with the warm chromosphere, the temperature of which\cite{OGorman2020,Dupree2020} is sufficiently high for dust sublimation\cite{Gupta2020}. 
Therefore, a dust clump that condenses near the photosphere may soon be sublimated by chromospheric heating, or, if the radiative acceleration is insufficient, it may fall back to the photosphere to be sublimated\cite{Cannon2021}. 
Considering these points, we conclude that the enhancements of $A(V)$ and $\tau _{10}$ during the Dimming may have occurred very close to the photosphere. 
This hypothesis is also supported by the fact that the polarized brightness of this circumstellar envelope changed significantly around the Dimming\cite{Cotton2020,Safonov2020}.

\subsection{Transition before the Dimming seen with \ce{H2O}}\label{ssec:EarlierTransition}

In order to further examine the dust-condensation process, we here focus on the time variation of the amount of molecular gas around Betelgeuse, in which the dust condensation occurred. 
We here focus on \ce{H2O} molecules. 
In addition to the dust emission seen in the wavelengths around $10\,\micron $, the amount of molecules, particularly \ce{H2O}, can also be traced using Himawari-8 photometric data. 
Since \ce{H2O} molecules in the Earth's atmosphere absorb starlight\cite{Shaw2013} in the wavelength ranges where the \ce{H2O} molecular features coming from stars exist\cite{Polyansky2018}, the observation of extraterrestrial \ce{H2O} molecules usually requires space telescopes\cite{Tsuji2000}. 
One of the missions of the Himawari-8 satellite is to observe the distribution of water vapor in the Earth's upper atmosphere\cite{Bessho2016}, and thus Himawari-8 has the wavelength bands where \ce{H2O} features exist. 
These bands, Bands~8--10 at around $6\text{--}8\,\micron $, are therefore useful for time-series observations of \ce{H2O} molecules around Betelgeuse. 
We focus here on the flux variation of Band~8, where the signal-to-noise ratio of the \ce{H2O} feature is maximized, thanks to the strength of this feature\cite{Perrin2007,Polyansky2018} and the sensitivity of the Himawari-8 satellite~(Extended Data \autoref{table:AHI}). 
We note that the $6\text{--}8\,\micron $ wavelength range is a complicated region in which other molecular bands are present, e.g., \ce{SiO} molecules. 
Thus, we should keep in mind the possibility of contamination by these molecules, and further spectroscopic observations are required to fully disentangle the problem. 
Nevertheless, Band~8 is the least contaminated by other molecules\cite{Tsuji2000,Perrin2007,Davies2021}. 

\begin{figure}[t!]
\includegraphics[width=\columnwidth ]{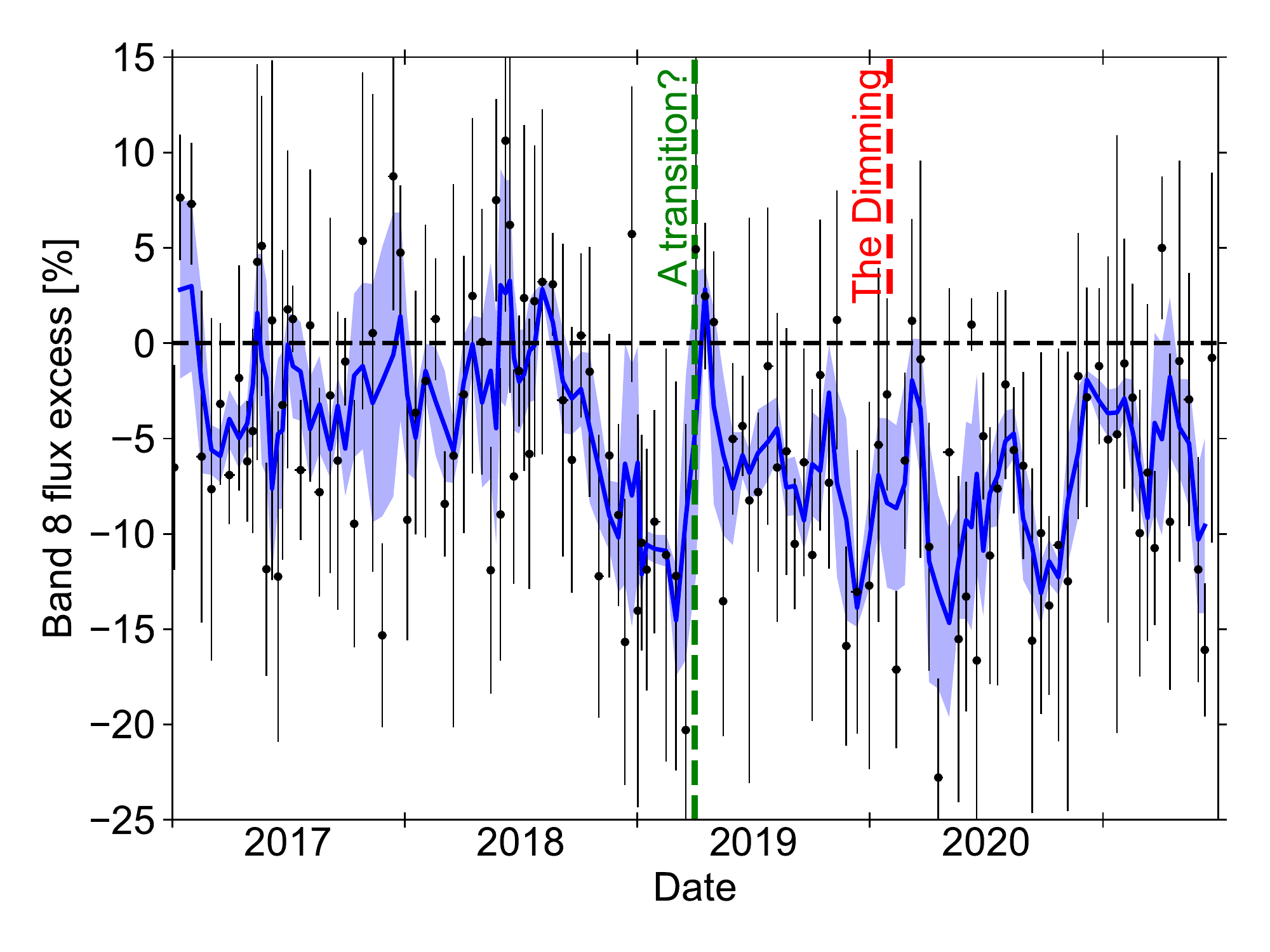}
\caption{\textbf{Molecular \ce{H2O} emission and absorption. } This figure shows the flux excess in Band~8 where molecular \ce{H2O} features exist. The blue line shows a smoothed version by averaging five points of the original black dots to guide the eye. We found a rapid transition in the flux excess from negative (absorption) to positive (emission) in early April 2019 (green vertical line), which occurs ten months before the Great Dimming in early February 2020 (red vertical line). }
\label{fig:B08residual}
\end{figure}

\autoref{fig:B08residual} shows the time variation of the flux excess in Band~8, defined as the ratio of the observed flux to predicted photospheric flux minus $1$, which is likely to trace the amount of \ce{H2O} molecular gas around Betelgeuse, maybe in the MOLsphere\cite{Tsuji2000,Perrin2007}.  
When gas clumps outside the photosphere are present that emit or absorb \ce{H2O} molecular features, the excess is positive or negative, respectively. 
In the first half of the observing period, i.e., before September 2018, to within the error bars the flux excess was almost consistent with zero, which indicates that there was no extra detectable \ce{H2O} gas around the photosphere. 
In contrast, most of the data points after September 2018 show negative flux excesses. 
This indicates the emergence of a clump and/or a layer of cool gas containing \ce{H2O} molecules, in which the dust that dimmed Betelgeuse may have formed. 
Considering the clumpy nature of the dust cloud that dimmed Betelgeuse\cite{Montarges2021} and the existence of the warm gas clump found from UV observations of the \ion{Mg}{ii} h and k lines around October 2019\cite{Dupree2020}, the \ce{H2O} gas that we found is also likely to be a clump. 

Finally, and more interestingly, we tentatively identified a rapid transition in the \ce{H2O} feature from absorption to emission in early April 2019~(the green vertical line in \autoref{fig:B08residual}). 
After visual inspection of the unbinned light curve for Band~8, we found that the time-scale of this transition is likely to be about a week or less.
This time-scale is much shorter than either the theoretical hydrodynamic time-scale of a few weeks to a few months\cite{Chiavassa2011b} or to previously observed time-scales of changes in the \ce{H2O} features seen in evolved red stars\cite{Matsuura2002,GRAVITY2021}. 
Because the gas clump cannot move outside of the line-of-sight to the photosphere on such a short time-scale, this transition indicates the occurrence of an episodic bursty event. 
One may speculate that the cause of this unusual event is related to another rare event, the Great Dimming; if so, it would be easy to imagine that such an unusual event could occur. 
One conceivable scenario involves a shock wave passing through the clump. 
It has been suggested that such a shock emerged from the bottom of the photosphere in January 2019 and propagated from there to the outer envelope of Betelgeuse\cite{Kravchenko2021}. 
Together with the fact that $A(V)$ began to increase at the time of this transition~(the fourth panel of \autoref{fig:Timeseries}) this shock might be related to the triggering process of the Great Dimming. 
Future theoretical investigations and time series of spatially resolved infrared observations would enable us to attack this further arising mystery. 

\bibliographystyle{naturemag}
\bibliography{Himawari_Betelgeuse}

\clearpage

\setcounter{table}{0}
\setcounter{figure}{2}
\renewcommand{\figurename}{Extended Data Fig.}
\renewcommand{\tablename}{Extended Data Fig.}

\begin{methods}
\subsection{}\vspace{-0.5cm}

In this paper, we first aim to establish and test a brand-new concept, using meteorological satellites as astronomical space telescopes, to obtain multi-band, long-term, and continuous light curves uninterrupted by the Sun. 
Second, we applied this technique to Betelgeuse to investigate its mysterious Great Dimming. 

Images of celestial objects, particularly the Moon, obtained by meteorological satellites have long been used to calibrate and ensure the quality of Earth observations\citemt{Kieffer1997,Fulbright2015,Stone2020}. 
In contrast, to our knowledge, this work is the first attempt to use these images to investigate stellar astrophysics. 
Here, we thoroughly describe the data properties, photometric methods, and their validation.

\subsection{Basic property of Himawari-8's scanning observation}

Geostationary meteorological satellites, such as Himawari-8, observe the entire disk of the Earth by sequential scans with several rows, each of which is termed a swath\cite{Bessho2016}\citemt{Kalluri2018}. 
Each swath is scanned with a 1D-like detector array from west to east. 
Due to this scanning procedure, the properties of images obtained by Himawari-8 have several differences compared to those obtained by typical astronomical observations using 2D devices. 
Nevertheless, we can use the reduced and distributed data, termed the L1B product, as usual astronomical images, provided that several caveats are taken into consideration. 

First, pixel navigation and data sampling are well controlled and regular, and their errors are usually smaller than $0.5\,\mathrm{pix}$ for Himawari-8\citemt{Tabata2016}. 
Thus, we can precisely transform the position in the image onto its location in the sky. 

Second, the fill factor of the pixels in the north-south direction is almost $100\%$, and the east-west sampling interval is smaller than the pixel size\citemt{Kalluri2018}. 
As such, we do not need to be concerned with undersampling due to gaps between pixels. 
Moreover, after the resampling process during the reduction described below, each image is transformed to the fixed square grid, which is very easy to use. 

Third, because each row of un-resampled images is observed by a single detector, inaccurate calibration of the detector occasionally introduces a stripe pattern into the obtained images. 
The strength of such a stripe has been shown to be small for observations of Earth scenes\citemt{Gunshor2020}, but it has not yet been closely examined for faint sources like stars. 
Therefore, it should be kept in mind that few flux measurements of stars are possibly influenced by this relatively large systematic error. 

Finally, other possibly minor effects include quantization noise, non-linearity of the detector response, and frequency of the calibration. 
We will further examine some causes of systematic and/or statistical errors and validate the photometric results in Supplementary Discussion. 

In this work, we limit our analyses to the images observed with the Advanced Himawari Imager~(AHI) onboard the Himawari-8 satellite, which is the first ``third-generation meteorological satellite'' characterized by access to as many as 16 bands\cite{Bessho2016}. 
There also are three other third-generation satellites equipped with 16-band imagers: GOES-16 and GOES-17\citemt{Schmit2005}, which are operated by the National Aeronautics and Space Administration~(NASA) and the National Oceanic and Atmospheric Administration~(NOAA), and GEO-KOMPSAT-2A\citemt{Kim2021}, which is operated by the Korea Meteorological Administration~(KMA). 
Although they have been operational since November 2016, March 2018, and July 2019, respectively, flux values outside the Earth's disk are found to be masked and unavailable in their published standard L1B products, which makes these satellites unusable for our analyses. 
Combined analyses with those satellites would improve the precision of our results, if their images of the space around the Earth's disk were to be made publically available. 

Reduction of the AHI images is performed by the Japan Meteorological Agency~(JMA)\citemt{Yokota2013,Kalluri2018}. 
This involves geometric re-gridding into a fixed coordinate system using a Lanczos-like resampling; a linear and global flux calibration using deep space as the background, with a solar diffuser for the optical and a blackbody for the infrared as standard flux sources; and nonlinear flux calibrations based on pre-launch ground tests. 
No corrections are made for atmospheric effects, nor is masking for cosmic rays performed.

\subsection{Photometry of the Himawari-8's image}\label{ssec:photometry}

About $6$-years of $16$-band images have been obtained by the Himawari-8 satellite at a frequency of once every ten minutes. 
We restricted the scope of our analyses to data observed from January 2017 to June 2021 because systematic biases in the flux calibration and in the positioning were identified in observations from July 2015 to November 2016~(see Supplementary Discussion). 
From this large number of images, we tried to find and measure the light from Betelgeuse, which is moving with the speed of the Earth's rotation, $15\,\mathrm{deg\,hour^{-1}}$, and which is found to be located in the field of view about once every two days. 
First, we calculated the times when Betelgeuse is in the direction of the Earth using the \textsc{Astropy} package\citemt{Astropy2013}, and we collected all the images in which Betelgeuse could be detected. 
Then we searched for the location of the Betelgeuse signal in each of these images in Bands~1--10, where the signal-to-noise ratio~(SNR) is high enough to detect the star. 
For each of those high-SNR bands, we searched for the location of the star by choosing the pixel with the largest counts among those pixels where Betelgeuse is expected to be located. 
In contrast, we inferred the locations in the other bands (Bands~11--16) by shifting from the position of Betelgeuse in Band~7, where the SNR is the highest, by around $30\,\mathrm{arcsec}$ depending upon the differences in the times of observation for the two bands. 

The pixel scale of the published images in each band is larger than the actual radius of the point-spread function~(PSF) estimated by the radius of the telescope. 
Thus, the light from a point source occupies only $1\times 1\,\mathrm{pix}^{2}$ of the un-reduced images, and it spreads out to ${\sim }2\times 2\,\mathrm{pix}^{2}$ during the reduction process (see the left panels of \autoref{fig:fulldisk/SEDalldate}). 
Due to this small PSF radius, PSF photometry does not work well, so instead we used the aperture-photometry technique with a $7\times 7\,\mathrm{pix}^{2}$ square aperture.
The background flux in the aperture was estimated and subtracted by fitting the flux within a square annulus around the star using a 2D polynomial function. 
We set the inner width of the annulus to be $7\,\mathrm{pix}$ and the outer width to be $43$, $21$, and $21\,\mathrm{pix}$ for the bands with the spatial resolutions of $0.5$, $1.0$, and $2.0\,\mathrm{km}$, respectively. 
We selected those widths in order to minimize the impact of stripe-shaped fluctuations of the background. 
We selected the order of the polynomial to be in the range $0\text{--}4$ for each measurement so that the Akaike Information Criterion calculated from the fitting residuals is minimized. 
Then, we estimated the stellar flux by summing the background-subtracted spectral irradiances within the $7\times 7\,\mathrm{pix}^{2}$ aperture. 

Since the background shapes are not guaranteed to be continuously connected between adjacent swaths, we excluded from our analyses cases in which the $7\times 7\,\mathrm{pix}^{2}$ aperture spans multiple swaths. 
Furthermore, we trimmed the square annulus used for background estimation by requiring that the annulus does not span multiple swaths.
Moreover, in order to avoid contamination from terrestrial lights, we introduced distance thresholds from the edge of the Earth's disk (see Extended Data \autoref{table:AHI}).

\subsection{Validation of the photometry}\label{ssec:validation_photometry}

Here, we performed a wide variety of validation steps, not only with Betelgeuse but also with other bright stars located in the field of view of the Himawari-8 satellite~(Extended Data \autoref{table:targets}). 
From the validation described in Supplementary Discussion and Supplementary Figs.~1--6, we found no evidence of any serious systematic bias in our photometry, and we thus conclude that the measurements are ready for further analyses. 
Nevertheless, it would be important, in a coming decade, to examine whether or not there are overlooked systematic effects in our method.

\subsection{SED fitting to the Himawari-8 light curves}\label{ssec:Model_photo}

The right panel of \autoref{fig:fulldisk/SEDalldate} shows the spectral energy distribution~(SED) averaged in time together with a model SED having the typical atmospheric parameters of Betelgeuse. 
The photometric bands with wavelengths $\lesssim 8\,\micron $~(Bands~1--10) are almost unaffected by the circumstellar dust emission, and thus they can be used to determine the atmospheric parameters of Betelgeuse. 
In contrast, the other bands receive radiation from silicate-dust emission at around $10\,\micron $, and they can thus be used to measure the amount of dust formed around Betelgeuse. 
We therefore first fitted the SED for Bands~1--10 with a model photosphere for Betelgeuse to determine the photospheric parameters---$R$, $T_{\mathrm{eff}}$, and $A(V)$---and we then determined $\tau _{10}$ using Bands~12--14, which have the highest SNR for the dust emission. 
Five examples of the fitting results are shown in Extended Data \autoref{fig:allbands_flux:Betelgeuse} \textbf{b}--\textbf{f}. 

Our final time-series photometric catalog of Betelgeuse is an incomplete one; the catalog lacks fluxes in several bands on several observations. 
This is especially serious for longer-wavelength bands, where the sky-background emission is stronger, and thus a larger number of photometric measurements was removed from the catalog when the star was located near the edge of the Earth's disk. 
Moreover, the signal-to-noise ratios in the mid-infrared bands are typically small, ${\sim }1$. 
In order to deal with these difficulties, we binned each light curve over $5\text{--}12$ observations requiring that all the binned data points for Bands~1--10 include five or more points. 
Note that the number of days in each bin is automatically determined by the observation frequency, which depends on the observation epoch. 
For example, there is a clump of observations in May--June 2018 with a frequency of $0.85\,\mathrm{days}$, which leads to the smaller number of days in the bins. 
Finally, we calculated the resulting median and its standard error for each bin.

\subsubsection{Model SED for the photosphere}

We fitted each point of this ``complete'' SED curve for Bands~1--10 with a photospheric model SED using the \textsc{LMFIT} package\citemt{Newville2014}. 
To obtain the model spectrum for the photosphere, we interpolated a grid of BT-NextGen model spectra\citemt{Allard2012} with solar metallicity and the surface gravity $\log g=0.0$, assuming that this one-dimensional model can be used to model the spectra of Betelgeuse at any phase. 
We used the Cardelli's reddening law\citemt{Cardelli1989} with the selective-to-total extinction ratio $R_{V}=3.1$\citemt{Levesque2005}, assuming that this law can be extrapolated to longer wavelengths. 
We also used the parallax of $6.55\,\mathrm{mas}$ measured with the \textsc{Hipparcos} satellite\citemt{vanLeeuwen2007}. 
Although the accuracy of the parallax of Betelgeuse is controversial\citemt{Harper2017,Joyce2020}, the choice of parallax value only changes the fitted radius in a relative way and does not affect the interpretation of the results. 
For the same reason, we do not take into account the error in the parallax value. 
The model spectra are then convolved with the response function of the AHI (September 2013 version retrieved from \url{https://www.data.jma.go.jp/mscweb/en/himawari89/space_segment/spsg_ahi.html}) to calculate the model SEDs for Betelgeuse, and we fitted the observed SED with the model.

\subsubsection{Dependence on the model selection}

For the grid of model spectra, we first tried the MARCS grid\citemt{Gustafsson2008} with $15M_{\odot }$\citemt{Ekstrom2012,Dolan2016,Wheeler2017,Joyce2020,Luo2022}, which is usually used to model the spectra of red supergiants\citemt{Levesque2005,Davies2013}. 
We found that the MARCS grid gives a smaller Akaike Information Criterion than the BT-NextGen grid, probably because the wavelengths and the strengths of the molecular bands in the MARCS grid are more accurate. 
Moreover, considering the fact that several previous works\citemt{Levesque2005} succeeded in reproducing observed optical spectra of red supergiants dominated by molecular \ce{TiO} bands with MARCS models, MARCS models appear to be an appropriate grid for the synthesis of \ce{TiO} bands in red supergiants. 
However, we found that the results of the fits obtained using the MARCS grid were physically unreasonable for most cases that we considered, e.g., $A(V)<0\,\mathrm{mag}$, although the reason, possibly related to 3D effects and/or circumstellar gas, is unclear. 
Thus, instead of the MARCS grid, we used the BT-NextGen grid to compute the model SED. 
Testing both the MARCS and Next-Gen grids, we found that the resulting model parameters we obtained using the BT-NextGen grid vary with time in the same way as those obtained using the MARCS grid; i.e., the differences between the time variations obtained using both grids are roughly constant~(see the left panels of Extended Data \autoref{fig:MARCS4par}). 
Therefore, the results of this paper using the Next-Gen grid are not affected quantitatively by the choice of model grid, as long as the time variations are interpreted in a relative way. 

We have also tested whether or not other model assumptions affect the results. 
For example, a change in $\log g$ from $0.0$ to $-0.5$ makes $T_{\mathrm{eff}}$ systematically ${\sim }80\,\mathrm{K}$ warmer and $A(V)$ systematically ${\sim }0.3\,\mathrm{mag}$ larger. 
Similarly, changing $R_{V}$ from $3.1$ to $4.4$\citemt{Levesque2005} decreases $T_{\mathrm{eff}}$ by $\lesssim 10\,\mathrm{K}$ and $A(V)$ by $\lesssim 0.05\,\mathrm{mag}$. 
From these tests, we concluded that although the absolute values of $R$, $T_{\mathrm{eff}}$, and $A(V)$ depend on the model assumptions, their relative variations appear to have been determined to the accuracy ${\sim }10\,\mathrm{K}$ in $T_{\mathrm{eff}}$ and ${\sim }0.05\,\mathrm{dex}$ in $A(V)$, both of which are much smaller than the fitting errors, ${\sim }40\,\mathrm{K}$ and ${\sim }0.2\,\mathrm{dex}$, respectively.

\subsubsection{Three-dimensional~(3D) effects in the photosphere}

The temperatures that previous works\cite{Levesque2020,Kravchenko2021}\citemt{Zacs2021,Alexeeva2021} and we have both derived assumed a one-dimensional~(1D) geometry for the model atmosphere. 
Thus these temperatures are affected by three-dimensional~(3D) effects in the optical spectrum of the real star\cite{Harper2020b}. 
In fact, the SED of a model with a given $T_{\mathrm{eff}}$, especially the strength of the \ce{TiO} molecular absorption bands that appear in the optical spectra of red supergiants\cite{Levesque2020}, has a variety of shapes due to 3D effects\cite{Chiavassa2011b}. 
Therefore, when we fit the observed SED of the real 3D photosphere of Betelgeuse with a 1D model SED, the fitted values of $T_{\mathrm{eff}}$ and $A(V)$ depend not only on $T_{\mathrm{eff}}$ and $A(V)$ themselves but also on the 3D effects\cite{Chiavassa2011b}. 
In particular, a time variation of the patchiness of the surface\citemt{LopezAriste2022,Aronson2022} could affect the time variations of $T_{\mathrm{eff}}$ and $A(V)$\cite{Harper2020b}. 

As a simplified test of the 3D effect on our fitting result, inspired by the optical image of Betelgeuse obtained during the Dimming\cite{Montarges2020,Montarges2021}, we considered a model photosphere composed of two regions with different $T_{\mathrm{eff}}$, which is similar to the model used by a previous work\cite{Harper2020b}. 
In this model, we assume that the cooler part of the photosphere has a varying temperature and a varying area and that the warmer part has a fixed temperature of $3650\,\mathrm{K}$. 
In contrast to the previous model\cite{Harper2020b}, we are able to consider two additional free parameters, $A(V)$ and $R$; without these two parameters we could not obtain a satisfactory fit to our multi-wavelength data. 
Altogether, we considered four free parameters: $R$, $A(V)$, and the temperature and area fraction of the cooler part; e.g., when the entire area of the photosphere is occupied by the cooler part, the fraction is $1.0$. 
With this four-parameter model, due to the strong degeneracy between $A(V)$ and the area, sometimes the fitting failed and gave unphysical results. 
To remove these cases, we excluded fitted results with areas greater than $1.0$ or lower than $0.0$ as well as results for which the temperatures reached the boundaries of the acceptable range (we assumed $2800$ and $4000\,\mathrm{K}$ for the lower and upper boundaries, respectively). 

The right panels of Extended Data \autoref{fig:MARCS4par} compare the fitted results obtained with the three-parameter model adopted in the main text and the four-parameter model described here. 
We found that with the four-parameter model the contribution of the increased $A(V)$ to the Dimming decreases but that it is still larger than zero (${\sim }0.4\,\mathrm{mag}$). 
This result supports the robustness of our results in a qualitative sense. 
Since the four-parameter model examined here is too simple to take 3D effects fully into account\cite{Harper2020b}, future theoretical modeling with self-consistent hydrodynamic models\citemt{Chiavassa2010} will be necessary.

\subsubsection{Comparison to previous radial-velocity (RV) measurements}

To check our SED fitting results, we compared the radius determined herein to previous radial-velocity (RV) measurements. 
Because the radius is given by the integration of the RV, we can validate the radius curve using literature RV curves. 

First, we found time lags of up to ${\sim }80\,\mathrm{days}$ between the times of maximum brightness and minimum radius (\autoref{fig:Timeseries}). 
Although such a large lag was not expected in a simple 1D hydrodynamic model of Betelgeuse\citemt{Joyce2020}, a lag between light and RV curves has been observationally found for Betelgeuse\cite{Dupree2020,Kravchenko2021}\citemt{Granzer2021}. 
Moreover, we found clockwise trajectories between the $R$/$T_{\mathrm{eff}}$ and Band-3 flux for most observation periods (Extended Data \autoref{fig:allbands_flux:Betelgeuse}\textbf{h}--\textbf{o}), which has been predicted theoretically\citemt{Hegger1997,Joyce2020}. 
The exception, the relation between $R$ and Band-3 flux between 27 October 2017 and 24 December 2018 shown in panel \textbf{i}, could be attributed to effects other than the fundamental-mode pulsation, such as the long-secondary period\citemt{Wasatonic2015}; however, the actual cause remains unclear. 

Second, we estimated the RV curve by calculating the differential of our radius curve, and compared it with literature RV curves\cite{Kravchenko2021}\citemt{Granzer2021}. 
This comparison reveals an overall good agreement between our data and the results cited above, supporting the validity of our radius measurement, and hence other parameters. 
However, minor differences of up to ${\sim }5\,\mathrm{km}\,\mathrm{s}^{-1}$ may exist, particularly in September 2018--November 2018 and September 2019--December 2019. 
Plausible causes of these differences are: (1) our imperfect modeling of Himawari-8's SEDs with a 1D model atmosphere, as discussed in the previous subsection, and (2) differences in the atmospheric layers that were observed in the present and previous studies, and hence the real, physical difference in RVs\cite{Kravchenko2021}.

\subsubsection{Model SED for the dust emission}\label{ssec:Model_dust}

After fitting the SED curve for Bands~1--10, we calculated the expected photospheric flux curves for Bands~11--16 and subtracted them from the observed flux curves. 
These residual fluxes are likely to come from the circumstellar dust emission, and they are useful for estimating the variation of the amount of circumstellar dust. 

For the model spectra of the circumstellar dust emission, we used the radiative-transfer code \textsc{DUSTY} V2\citemt{Ivezic1999}. 
For the settings of the \textsc{DUSTY} calculation, we basically followed the procedure described in a previous work\citemt{Beasor2016}, except for several modifications. 
First, we only fitted $\tau _{10}$ and fixed the other parameters. 
We used the ``warm silicates'' dust composition\citemt{Ossenkopf1992}, a constant grain size of $0.3\,\micron $, and a constant dust temperature of $1300\,\mathrm{K}$ at the inner boundary, which roughly corresponds to the sublimation temperature of silicate dust\cite{Gupta2020}. 
Although inner-boundary temperatures cooler than $1300\,\mathrm{K}$ are sometimes found by observations of red supergiants\citemt{Beasor2016}, we adopted this value since a different inner-boundary temperature only results in a systematic bias in $\tau _{10}$ as long as the temperature is independent of time. 
We also tried the assumption of a constant radius at the inner boundary, and we found that the results do not change in a qualitative sense. 
For the stellar radiation that illuminates the dust, we used the time variation of $T_{\mathrm{eff}}$ determined in the previous section to calculate the natal photospheric SED, after applying spline smoothing to reduce the noise in each measurement. 
Then, we calculated the expected silicate-dust emission in Bands~12--14 on the $\tau _{10}$ grid and determined the $\tau _{10}$ value that best matches the observed flux-residual curve for each of Bands~12--14. 
Finally, we calculated the mean of the three $\tau _{10}$ values determined from Bands~12--14 to obtain the final estimate of $\tau _{10}$. 
We note that $\tau _{10}$ that we determined is in fact not the actual optical depth at $10\,\micron $; it is rather a model parameter that we use to match the observed and model fluxes of Bands~12--14, and we use it as an indicator of the amount of circumstellar silicate around Betelgeuse.

\subsection{Code Availability}
We used the \textsc{LMFIT} package\citemt{Newville2014} to fit the observed SED in the optical to near infrared with publicly available NextGen\citemt{Allard2012} and MARCS\citemt{Gustafsson2008} model grids. 
We also used the \textsc{DUSTY} V2 code\citemt{Ivezic1999} to calculate the dust emission. 
We have not made the codes for the photometry publicly available because they are not prepared for open use. 
Instead, we have made the photometric catalog available, as described in the Data Availability statement. 

\subsection{Data Availability}
Reduced images taken by the Himawari-8 satellite are publicly available at the Data Integration and Analysis System \textit{via} \url{https://diasjp.net/en/}. 
The final photometric product of this paper is available at \url{https://d-taniguchi-astro.github.io/homepage/Data_Himawari_en.html}, which is archived as Supplementary Data in Supplementary Information.


\bibliographystylemt{naturemag}
\bibliographymt{Himawari_Betelgeuse}

\begin{addendum}
\item[Correspondence and requests for materials]
should be addressed to Daisuke Taniguchi. 

\item[Acknowledgements] 
We thank Noriyuki Matsunaga, Takafumi Kamizuka, Kengo Tachibana, Takashi Onaka, Naoto Kobayashi, Mingjie Jian, and Makoto Ando for comments and discussion. 
We also thank the Meteorological Satellite Center of JMA for explanations of the details of the Himawari-8 observations, in response to our inquiries. 
This work has been supported by Masason Foundation. 
D.T., K.Y., and S.U. are financially supported by JSPS Research Fellowship for Young Scientists and accompanying Grants-in-Aid for JSPS Fellows~(Nos.~21J11555, 21J11266, and 21J20742, respectively). 
D.T. also acknowledges financial support from Masason Foundation since 2018, The University of Tokyo Toyota-Dwango Scholarship for Advanced AI Talents in 2020--21, and from Iwadare Scholarship Foundation in 2020. 
S.U. is supported by the WINGS-FoPM program of the University of Tokyo. 
This study used the Himawari data downloaded from the Data Integration and Analysis System (DIAS) by the University of Tokyo. 
This study also used the variable-star observations from the AAVSO International Database contributed by observers worldwide. 
Data analysis was in part carried out on the Multi-wavelength Data Analysis System operated by the Astronomy Data Center (ADC) of National Astronomical Observatory of Japan, and in a computer environment funded by JSPS KAKENHI (Nos.~JP25287119 and 20H05731).

\item[Author contributions]
D.T. and S.U. initiated this work. 
D.T. analyzed the photometric datasets. 
K.Y. led the photometry of the Himawari-8 images, and D.T. and S.U. partly contributed to it. 
All the authors contributed substantially to the discussion and the writing. 

\item[Competing interests]
The authors declare no competing interests. 

\item[Additional Information]
\item[Peer review information]
\item[Reprints and permissions information]
is available at \url{http://www.nature.com/reprints}. 

\end{addendum}

\end{methods}

\begin{extendeddata}

\begin{table*}
\centering 
\begin{threeparttable}[t!]
\caption{\textbf{Basic properties of the Himawari-8 AHI images and the thresholds used for photometry. }}
\label{table:AHI}
\begin{tabular}{cSSSSSSSSSS} \toprule 
Band & \multicolumn{1}{c}{$\lambda _{0}$\tnote{a}} & \multicolumn{1}{c}{$\Delta \lambda $\tnote{a}} & \multicolumn{2}{c}{Pixel scale} & \multicolumn{1}{c}{$m_{\mathrm{AB}}^{}$\tnote{c}} & \multicolumn{1}{c}{$F_{\mathrm{max}}$\tnote{d}} & \multicolumn{1}{c}{LSB\tnote{e}} & \multicolumn{1}{c}{$z_{\mathrm{min}}$\tnote{f}} & \multicolumn{2}{c}{LB test $p$ value\tnote{g}} \\
No. & \multicolumn{1}{c}{(\micron )} & \multicolumn{1}{c}{(\micron )} & \multicolumn{1}{c}{(km)\tnote{b}} & \multicolumn{1}{c}{(arcsec)} & \multicolumn{1}{c}{(mag)} & \multicolumn{2}{c}{($\mathrm{nW}\,\mathrm{m}^{-2}\,\micron ^{-1}$)} & \multicolumn{1}{c}{(km)} & \multicolumn{1}{c}{Rigel} & \multicolumn{1}{c}{Procyon} \\ \midrule 
1 & 0.471 & 0.041 & 1.0 & 5.76 & 2.4 & 597 & 0.38 & 75 & 0.419 & 0.509 \\
2 & 0.510 & 0.031 & 1.0 & 5.76 & 2.5 & 560 & 0.35 & 75 & 0.174 & 0.210 \\
3 & 0.639 & 0.082 & 0.5 & 2.88 & 2.4 & 121 & 0.31 & 75 & 0.003 & 0.321 \\
4 & 0.857 & 0.034 & 1.0 & 5.76 & 0.8 & 288 & 0.182 & 60 & 0.884 & 0.859 \\
5 & 1.610 & 0.041 & 2.0 & 11.53 & 0.7 & 287 & 0.045 & 65 & 0.341 & 0.204 \\
6 & 2.257 & 0.044 & 2.0 & 11.53 & 0.9 & 89.1 & 0.0141 & 60 & 0.938 & 0.609 \\
7 & 3.89 & 0.20 & 2.0 & 11.53 & 2.0 & 40.2 & 0.00079 & 55 & 0.964 & 0.329 \\
8 & 6.24 & 0.82 & 2.0 & 11.53 & 0.3 & 29.9 & 0.0048 & 90 & 0.331 & 0.836 \\
9 & 6.94 & 0.40 & 2.0 & 11.53 & -0.1 & 36.4 & 0.0058 & 100 & 0.288 & 0.396 \\
10 & 7.35 & 0.19 & 2.0 & 11.53 & -0.3 & 38.8 & 0.0031 & 100 & 0.893 & 0.382 \\
11 & 8.59 & 0.37 & 2.0 & 11.53 & -0.7 & 50.6 & 0.0040 & 100 & 0.270 & 0.500 \\
12 & 9.64 & 0.38 & 2.0 & 11.53 & -0.8 & 49.7 & 0.0039 & 130 & 0.689 & 0.316 \\
13 & 10.41 & 0.42 & 2.0 & 11.53 & -1.2 & 47.5 & 0.0038 & 90 & 0.756 & 0.870 \\
14 & 11.24 & 0.67 & 2.0 & 11.53 & -1.0 & 44.3 & 0.0035 & 80 & 0.913 & 0.250 \\
15 & 12.38 & 0.97 & 2.0 & 11.53 & -1.4 & 39.4 & 0.0031 & 100 & 0.658 & 0.285 \\
16 & 13.28 & 0.56 & 2.0 & 11.53 & -2.5 & 36.0 & 0.0058 & 100 & 0.351 & 0.074 \\ \bottomrule 
\end{tabular}
\begin{tablenotes}[flushleft,normal]
\item[a] Each band is centered on $\lambda _{0}$ and has the FWHM $\Delta \lambda $. 
\item[b] Spatial resolution at the sub-satellite point, i.e., $35,786\,\mathrm{km}$ from the satellite. 
\item[c] Sky-$3\sigma $ limiting AB magnitude for each band at the declination of Betelgeuse. 
\item[d] Maximum valid flux per pixel in the calibrated L1B file format of Himawari-8 images. According to results of ground tests\citemt{Okuyama2018}, the saturation flux of the AHI itself is larger than these values. 
\item[e] Least significant bit (LSB) of L1B images; i.e. the flux of each pixel is measured in increments of LSB. 
\item[f] Minimum distance between the Betelgeuse line of sight and the edge of the Earth's edge accepted for photometric analyses in this study. 
\item[g] $p$ value of Ljung-Box test\citemt{Ljung1978}, with the time delays up to $30\,\mathrm{days}$ for the two non-variable stars, Rigel and Procyon. 
\end{tablenotes}
\end{threeparttable}
\end{table*}

\begin{table*}
\centering 
\begin{threeparttable}[t!]
\caption{\textbf{The stars for which we performed photometry. } }
\label{table:targets}
\begin{tabular}{lrlcc} \toprule 
\multicolumn{1}{c}{Name} & \multicolumn{1}{c}{HD} & \multicolumn{1}{c}{Sp. Type\tnote{a}} & Dec.\tnote{a} & Frequency\tnote{b} \\
 & \multicolumn{1}{c}{number} & & (deg) & ($\mathrm{days}^{-1}$) \\\midrule 
Betelgeuse & 39801 & M1--M2Ia--Iab & $+7.41$ & $0.58$ \\
Rigel & 34085 & B8Iae & $-8.20$ & $0.91$ \\
Procyon & 61421 & F5IV--V+DQZ & $+5.22$ & $0.33$ \\
$\delta $~Oph & 146051 & M0.5III & $-3.69$ & $0.29$ \\
$\alpha $~Cet & 18884 & M1.5IIIa & $+4.09$ & $0.48$ \\ \bottomrule 
\end{tabular}
\begin{tablenotes}[flushleft,normal]
\item[a] Spectral type and declination (J2000.0) taken from the SIMBAD database\citemt{Wenger2000} on 15 June 2021. 
\item[b] Typical frequency at which the star appear in the Himawari-8 images. 
\end{tablenotes}
\end{threeparttable}
\end{table*}

\begin{figure*}[t!]
\includegraphics[width=2\columnwidth ]{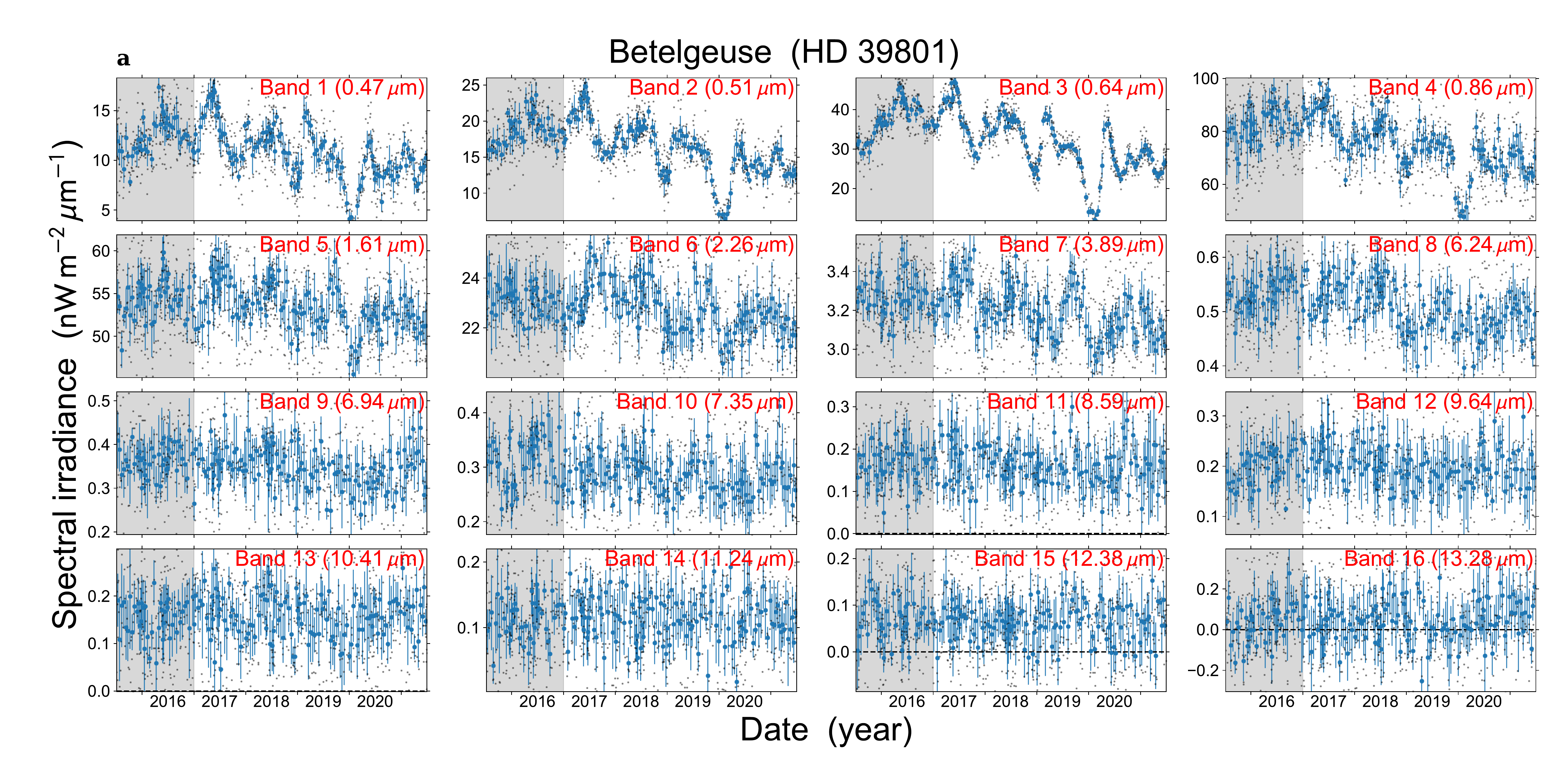}
\includegraphics[width=2\columnwidth ]{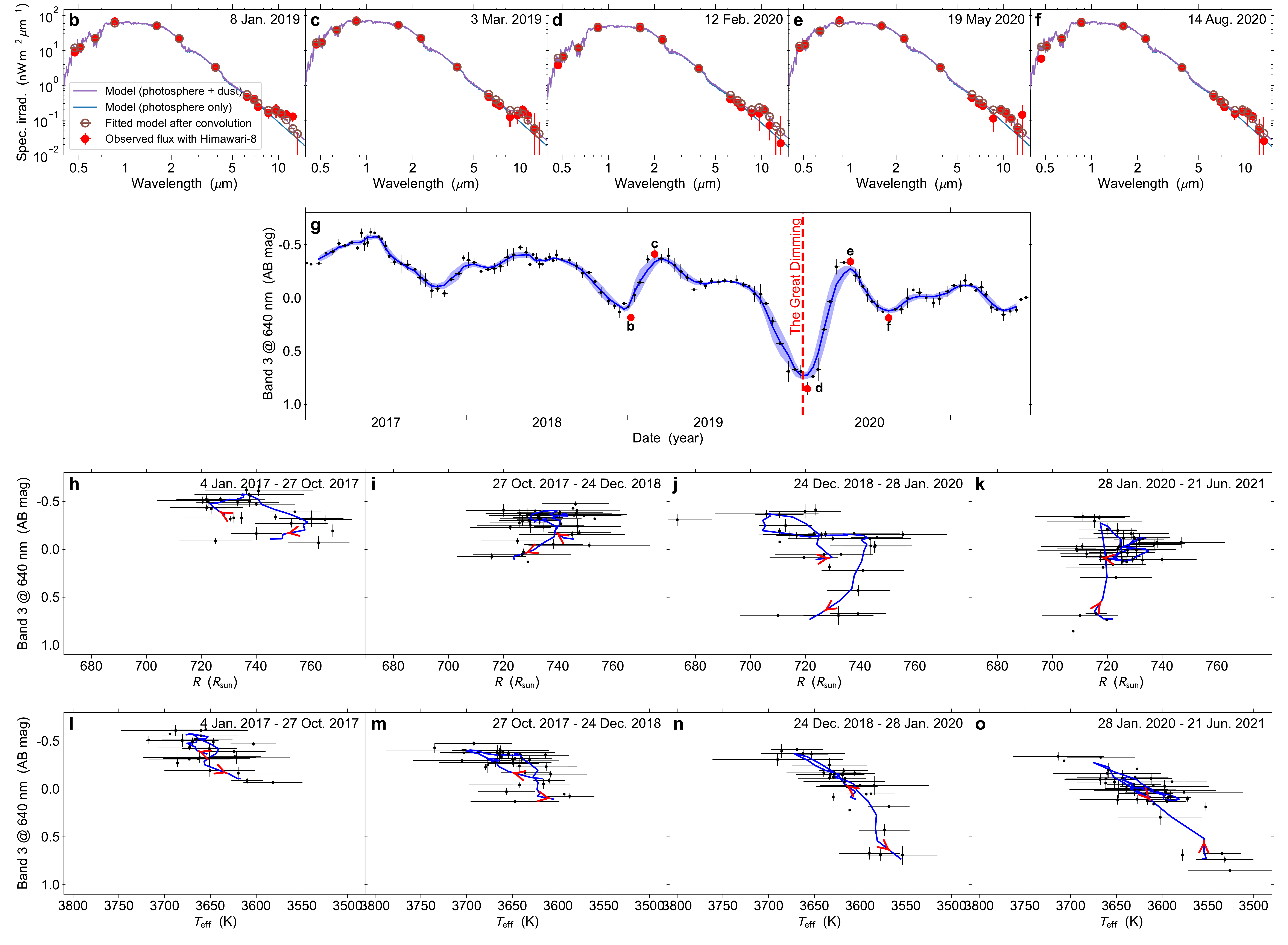} \vspace{-0.35cm}
\caption{\textbf{Results of the photometry and the SED fitting for Betelgeuse. } \textbf{a}, Light curves of Betelgeuse in all bands. The black dots represent individual measurements, and the blue circles and vertical bars represent the medians and standard errors after binning, respectively. The photometric measurements in the gray-shaded areas, i.e., in 2015 and 2016, were not used in the analysis. \textbf{b}--\textbf{f}, Five results of the SED fitting as examples, whose observation epoch are shown in \textbf{g}. The red and brown circles represent the observed and fitted spectral irradiances for each band, respectively. The purple and blue lines represent the fitted spectrum before the convolution with the response function. \textbf{g}, The light curve of Betelgeuse at Band~3 in AB magnitudes, as in \autoref{fig:Timeseries}. \textbf{h}--\textbf{k} and \textbf{l}--\textbf{o}, The relation between the radius $R$ or effective temperature $T_{\mathrm{eff}}$, respectively, and Band~3 magnitude with the smoothed trajectory created by averaging five points. }
\label{fig:allbands_flux:Betelgeuse}
\end{figure*}

\begin{figure*}[t!]
\centering 
\includegraphics[width=2\columnwidth ]{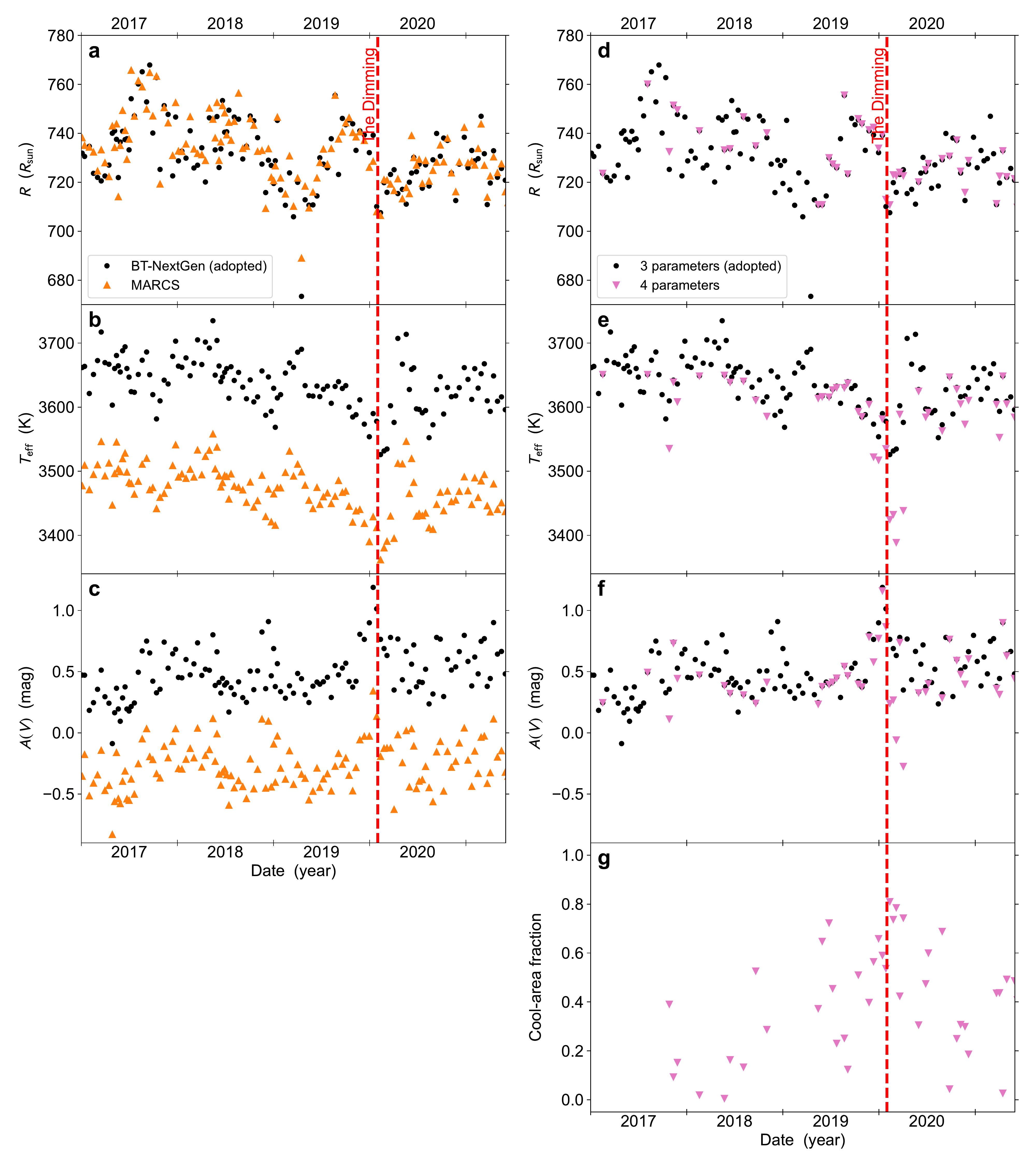}
\caption{\textbf{Comparison of the results of SED fitting with different model assumptions. } In all the panels, the black circles indicate the results adopted in this paper (the same as \autoref{fig:Timeseries}), which use the grid of Next-Gen model spectra\protect\citemt{Allard2012}, assuming three free parameters in the fitting. \textbf{a}--\textbf{c}, The effect of the adopted model-spectrum grid: Next-Gen\protect\citemt{Allard2012}~(blue circles) \textit{vs.} MARCS\protect\citemt{Gustafsson2008}~(upward-pointing orange triangles). \textbf{d}--\textbf{g}, The effect of the number of parameters: three parameters~(blue circles) \textit{vs.} four parameters~(downward-pointing pink triangles). }
\label{fig:MARCS4par}
\end{figure*}

\end{extendeddata}

\clearpage

\setcounter{page}{1}
\setcounter{table}{0}
\setcounter{figure}{0}
\makeatletter
\renewcommand{\figurename}{Supplementary Fig.}
\renewcommand{\tablename}{Supplementary Table}
\makeatother

\begin{supplementaryinformation}
\subsection{}\vspace{-0.5cm}

In this Supplementary Information pdf, we describe our validation steps of the photometry. 

\subsection{Signal-to-noise ratio (SNR) of the imager}

According to the results of the pre-launch verification carried out by JMA\citesi{Okuyama2018_si}, the noise level of AHI is around $0.7$, $0.2$, and $0.02\,\mathrm{nW}\,\mathrm{m}^{-2}\,\micron ^{-1}$ per pixel for visible, near-infrared, and mid-infrared bands, respectively. 
Assuming independence of noise among pixels, the SNR of our flux measurements for Betelgeuse due to noise would be approximately $2\text{--}20$ and $0.1\text{--}1$ for Bands~1--10 and 11--16, respectively, before applying the temporal binning described later. 
Those estimated SNRs are roughly consistent with the sample-to-sample variation of our photometric results (Extended Data \autoref{fig:allbands_flux:Betelgeuse}\textbf{a}).

\subsection{Calibration accuracy}
Long-term in-orbit monitoring of the AHI  (\url{https://www.data.jma.go.jp/mscweb/en/oper/eventlog/Update_of_Calibration_Information_2020.pdf}) indicates that the sensor gain in the visible and near-infrared bands (Bands~1--6) is affected by seasonal fluctuations of up to $0.5\%$ and by a quasi-linear fading trend of up to $0.5\%\,\mathrm{year}^{-1}$.
The estimated stellar parameters are affected by the uncertainties due to these calibration biases. 
However, the impact of the biases is likely to be negligible, since these are sufficiently smaller than the random ${\sim }5\%$ errors in the observed binned fluxes of Betelgeuse in the visible bands (Extended Data \autoref{fig:allbands_flux:Betelgeuse}\textbf{a}).
Moreover, the weakening trend results only in a weak, long-term monotonic trend in the stellar parameters of Betelgeuse, which does not affect the interpretation of the results.

\subsection{Saturation}
Since the AHI has been designed to observe the Earth, it is necessary to confirm that stellar observations are not affected by the saturation of the detector. 
We confirmed that the maximum valid fluxes of the AHI in all bands (Extended Data \autoref{table:AHI}) are sufficiently larger than the observed fluxes of Betelgeuse shown in the right panel of \autoref{fig:fulldisk/SEDalldate}. 
Therefore, the results of this study are not affected the saturation.

\subsection{Quantization noise}
In the opposite direction to saturation, the photometry of AHI data might be affected by quantization during the reduction. 
Due to the limited size of data that can be downlinked in real time, satellite data is compressed to 11 to 14 bits\citesi{Kalluri2018_si}. 
Thus, the quantization noise caused by the relatively large least significant bit (LSB) of each band (Extended Data \autoref{table:AHI}) would have an impact on the photometric error. 
However, we confirmed that the quantization noise is at least two times smaller than the photometric error due to background fluctuation in all bands. 
Moreover, with the exception of Band~16, which we did not use for the analysis of Betelgeuse, the typical flux of Betelgeuse is at least five times larger than the quantization noise. 
This consideration demonstrate that the quantization noise has a non-negligible, but non-dominant contribution to the photometric error.

\subsection{Point-spread function (PSF)}
The PSFs of the reduced Himawari-8 images were estimated by stacking the flux of Betelgeuse around the brightest pixel. 
The results in Supplementary \autoref{fig:PSF} show that the PSF occupies ${\sim }2\times 2\,\mathrm{pix}^{2}$, regardless of the wavelength or the pixel scale of the observation band; i.e., $0.5\,\mathrm{km}$ for Band~3, $1.0\,\mathrm{km}$ for Bands~1, 2, and 4, and $2.0\,\mathrm{km}$ for the others. 
This finding supports the conclusion that the PSF mainly results from the resampling process in the reduction. 
Because of the method used for resampling\citesi{Kalluri2018_si}, we can safely assume that the signal from a point source like a star does not spread further than $\pm 2\,\mathrm{pix}$ away from the brightest pixel.

\subsection{Positioning accuracy}

Since the location of Betelgeuse in the images of the mid-infrared bands was inferred from that of Band~7, large errors in the inferred position, if present, would cause flux loss. 
To evaluate the accuracy of the inter-band positioning inference, we estimated the position of Betelgeuse in the visible and near-infrared bands from Band~7 in the same manner as for the mid-infrared bands, and we compared the inferred and actual positions~(Supplementary \autoref{fig:Position_accuracy}). 
Focusing on the period after January 2017, both in the east--west ($x$) and north--south ($y$) directions, we found that the positioning-inference errors are typically less than $5\,\mathrm{arcsec}$, which corresponds to $0.5\,\mathrm{pix}$ in the mid-infrared bands. 
Thus, our aperture size, $7\times 7\,\mathrm{pix}^{2}$, is large enough to cover all of the size of the circumstellar envelope of Betelgeuse ($<0.5\,\mathrm{pix}$\cite{Kervella2011}), the PSF ($\pm 2\,\mathrm{pix}$ at most), and the positioning error ($\pm 0.5\,\mathrm{pix}$ at most) after 2017. 
Therefore, it is unlikely that the results in the mid-infrared bands missed flux from Betelgeuse due to PSF and/or positioning errors after January 2017. 
In 2015 and 2016, however, the positioning errors were considerably larger in some cases, especially in Band~5 (the middle panel of Supplementary \autoref{fig:Position_accuracy}). 
Thus, for safety, we excluded all measurements in 2015 and 2016 from our analyses.

\subsection{Light curves of bright stars other than Betelgeuse}

In order to validate the photometric results, we applied the photometric method we used for Betelgeuse to four bright stars within the field of view of the AHI~(Extended Data \autoref{table:targets}). 
Extended data \autoref{fig:allbands_flux:Betelgeuse}\textbf{a} and Supplementary Figs.~\ref{fig:allbands_flux:Rigel-Procyon}, \ref{fig:allbands_flux:delOph-alfCet} show the resulting $16$-band light curves for Betelgeuse, two non-variable stars (Rigel and Procyon), and two red giant stars ($\delta $~Oph and $\alpha $~Cet), using the observational data from the AHI in its period of operation from July 2015 to June 2021. 
The light curves of the two non-variable stars show that the fluxes in Bands~1--4 before the beginning of 2016 were systematically enhanced relative to the period after this. 
Similarly, the light curves of the two red giant stars also show enhancement in the early epoch. 
Though the cause of this enhancement, which could be related to an update in the reduction process by JMA in 9 May 2016\citesi{Tabata2016_si}, is unclear, in order to avoid this enhancement, we decided to use the data only after 1 January 2017 for the subsequent analyses. 

In order to test quantitatively whether there is time variability in the light curves of the two non-variable stars after January 2017, we tested the null hypothesis that there is no autocorrelation in the light curves with time delays up to $30\,\mathrm{days}$ using the Ljung--Box test\citesi{Ljung1978_si} implemented in the \textsc{statsmodels} package\citesi{Seabold2010_si}. 
We found that, among the $16$ bands for each of the $2$ stars, only one band of one star has a $p$ value smaller than $0.05$~(Extended Data \autoref{table:AHI}); i.e., there is no evidence of any detectable autocorrelation in the light curves of the non-variable stars. 
Considering that these two stars are sufficiently bright in the optical and near-infrared bands (Bands~1--5), this result indicates that our photometry in these bands is not likely to be affected by time-dependent systematic biases. 
Further, it indicates that photometry in the mid-infrared bands is always likely to have accurate zero points, and hence partially supports the reliability of the mid-infrared photometry. 
Nevertheless, it would be better to test the mid-infrared photometry with infrared-bright, non-variable stars; unfortunately, there are no such stars in the field of view of Himawari-8.

\subsection{Atmospheric contamination}
The radiance of a star measured while it is located close to the edge of the Earth's disk might be affected by emission and Rayleigh scattering from the terrestrial atmosphere, and thus these effects need to be examined. 
First, in order to evaluate the amount of spurious flux resulting from such contaminating emissions, we applied the photometric procedure used for Betelgeuse to empty points around the Earth, rather than to Betelgeuse. 
The measured spurious fluxes of the empty points correspond to the amount of bias in the photometry due to unsatisfactory background subtraction. 
As shown in the left panels of Supplementary \autoref{fig:ray_height}, the amount of background bias actually depends on the distance from the edge of the Earth's disk, and the dependence of this atmospheric emission differs among the bands. 
We confirmed that the measurements contaminated by this bias had been removed from our dataset after setting the thresholds for the Earth--Betelgeuse distances for the observed images in the individual bands~(Extended Data \autoref{table:AHI}). 

Next, we evaluated the effects of the terrestrial sky on starlight using the photometric data for the two non-variable stars, Rigel and Procyon, measured with the same distance thresholds. 
We found that the measured fluxes of the two stars have no significant dependence on the Earth--star distances (the right panels of Supplementary \autoref{fig:ray_height}). 
This fact indicates that our photometric measurements also are not likely to be affected by Rayleigh scattering or by other effects relating to the terrestrial atmosphere.

\subsection{Contamination by Solar-System objects}

Solar-System objects like the Moon and Venus sometimes appear in images from meteorological satellites\cite{Bessho2016}. 
If these objects are located too close to the stars whose fluxes are to be measured (\autoref{table:targets}), they could affect photometry. 
Using the \textsc{Astropy} package\citesi{Astropy2013_si}, we calculated the direction of bright Solar-System objects as seen from Himawari-8 when the stars are located within the field of view. 
Herein, we only considered bright objects that could be detected with Himawari-8: the Moon, Mercury, Venus, Mars, Jupiter, and Saturn. 
As a result, we confirmed that none of these objects approach too close to the stars, and thus their resulting contamination does not affect the photometry. 
We note that no other bright stars are present around each of the target stars. 

Another source of contamination is scattered light in the optical and near-infrared wavelengths from the Sun. 
This effect becomes strongest at around midnight when the satellite observes the direction close to the Sun, particularly around the equinoxes\citesi{Gunshor2020_si}. 
However, due to a close relationship between the observing month and hour for each star, midnights correspond to specific month(s), e.g., January and February for Betelgeuse. 
Similarly, we confirmed that none of our target stars are observed at around midnight during equinox periods. 
Moreover, we confirmed that scattered light was significantly reduced by the background-subtraction procedure. 
Although the background flux is as much as, e.g., $1.0$ and $0.6\,\mathrm{nW}\,\mathrm{m}^{-2}\,\micron ^{-1}$ per pixel for Bands~1 and 6, respectively, before the subtraction, they are reduced to $0.02$ and $0.003\,\mathrm{nW}\,\mathrm{m}^{-2}\,\micron ^{-1}$, respectively.

\subsection{Comparison with literature photometric data of Betelgeuse and Rigel}

From the validation steps so far, our light curves appear to be independent of both the date of observation and the distance from the edge of the Earth's disk; i.e., they seem to be accurate in a relative sense. 
Finally, in order to evaluate the absolute accuracy of the photometry, we compared the Himawari-8's photometric data for Betelgeuse and Rigel with literature measurements. 

First, the top panel of Supplementary \autoref{fig:SED:Betelgeuse-Rigel} compares the optical light curves obtained with Himawari-8's Bands~2 and 3 (centered at $0.51$ and $0.64\,\micron $, respectively) to the \textit{V}-band light curve ($0.55\,\micron $) cataloged in several sources: the American Association of Variable Star Observers~(AAVSO) database, a work doing daylight observation\citesi{Nickel2021_si}, and a work using images obtained by the STEREO Mission\citesi{Dupree2020b_si}. 
We found that our light curves are well parallel to that of the \textit{V} band. 
Moreover, as expected from the central wavelengths, the \textit{V}-band light curve is located between those of Bands~2 and 3 and is relatively close to Band~2. 

Second, the bottom panels of Supplementary \autoref{fig:SED:Betelgeuse-Rigel} compare Himawari-8 and literature SEDs of Betelgeuse and Rigel: the infrared spectra of Betelgeuse and Rigel taken with the Short-Wavelength Spectrometer (SWS) on the Infrared Space Observatory (ISO) satellite\citesi{Sloan2003_si} (TDT numbers~69201980 and 83301505, respectively); the mean spectral irradiances of Betelgeuse between January 2017 and June 2021 in the six visible and near-infrared wavebands~(\textit{B}, \textit{V}, \textit{R}, \textit{I}, \textit{J}, and \textit{H}) taken from the AAVSO database; and the spectral irradiances of Rigel in the same six bands\citesi{Ducati2002_si}. 
The figures show overall good agreement between the Himawari-8 SEDs and those from the literature, although there is a discrepancy between the two in the optical and near-infrared SED, especially for Betelgeuse in the near infrared. 
The reason for this discrepancy is unclear, but differences in the dates of observation and in the bandwidths may contribute to the discrepancy, at least in part. 

Considering these results, the absolute accuracy of the photometry is estimated conservatively as ${\sim }20\%$, although the absolute accuracy of the photometry does not affect the final results of this paper; i.e., the time variation of the brightness and stellar parameters of Betelgeuse.

\bibliographystylesi{naturemag}
\bibliographysi{Himawari_Betelgeuse}

\begin{figure*}[t!]
\centering 
\includegraphics[width=1.5\columnwidth ]{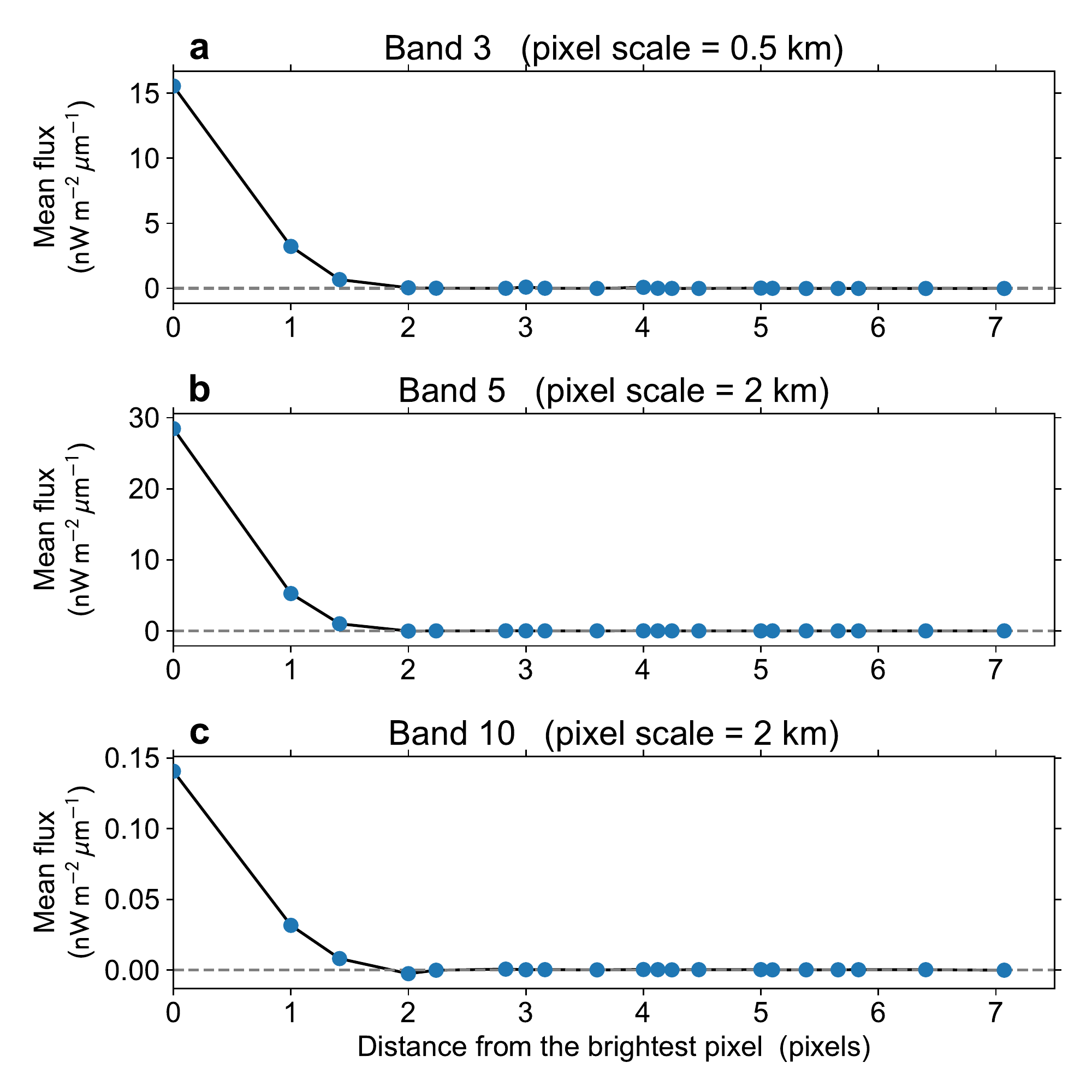}
\caption{\textbf{Point-spread function~(PSF). } The PSFs estimated using multiple images of Betelgeuse. \textbf{a}, Band~3. \textbf{b}, Band~5. \textbf{c}, Band~10.}
\label{fig:PSF}
\end{figure*}

\begin{figure*}[t!]
\centering 
\includegraphics[width=1.5\columnwidth ]{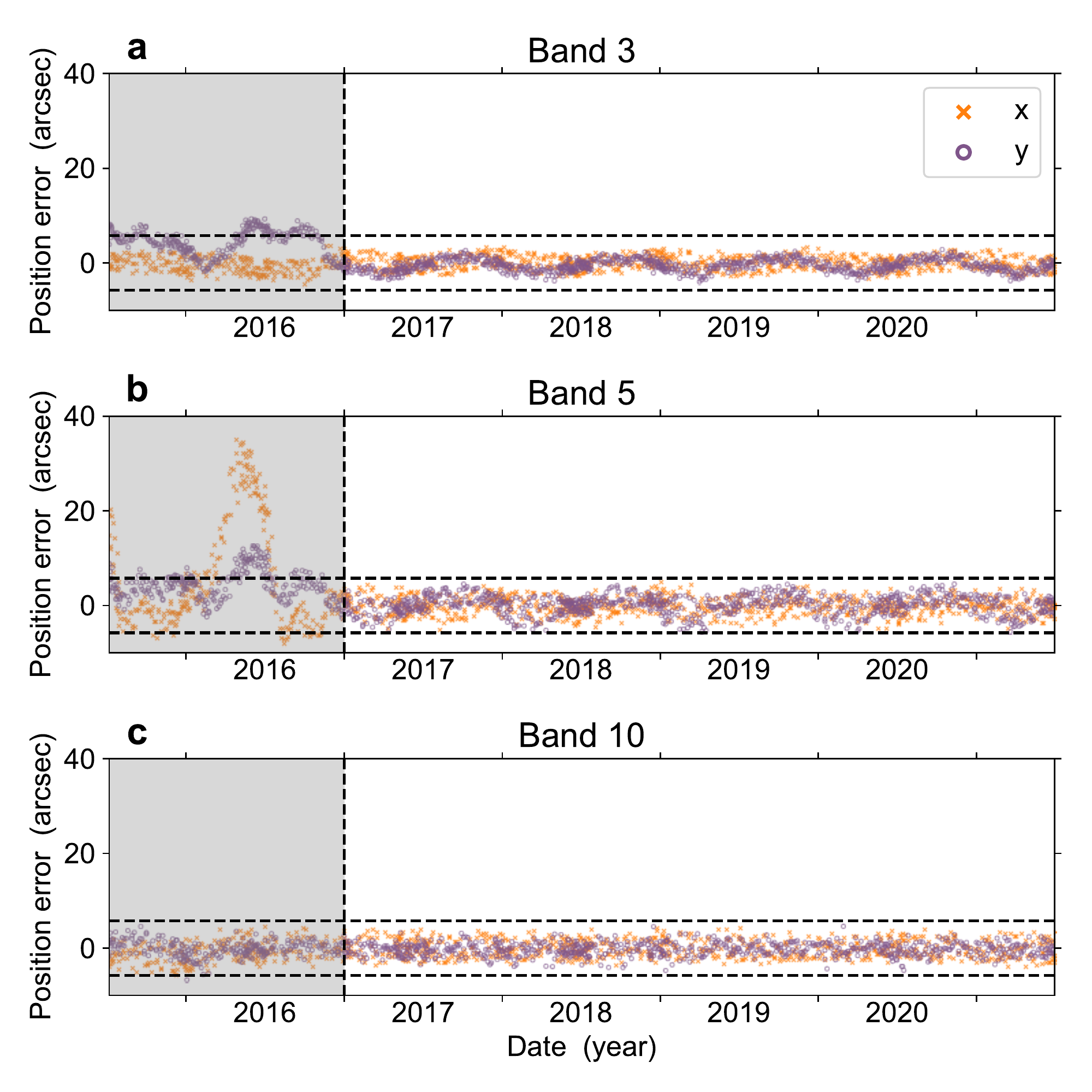}
\caption{\textbf{Positioning errors. } The errors of the position inferences. The broken lines at $\pm 11.53/2\,\mathrm{arcsec}$ represent the half-pixel scale of the mid-infrared bands. The orange crosses and the purple circles represent the errors in the $x$ (east--west) and $y$ (north--south) directions. \textbf{a}, Band~3. \textbf{b}, Band~5. \textbf{c}, Band~10.}
\label{fig:Position_accuracy}
\end{figure*}

\begin{figure*}[t!]
\includegraphics[width=2\columnwidth]{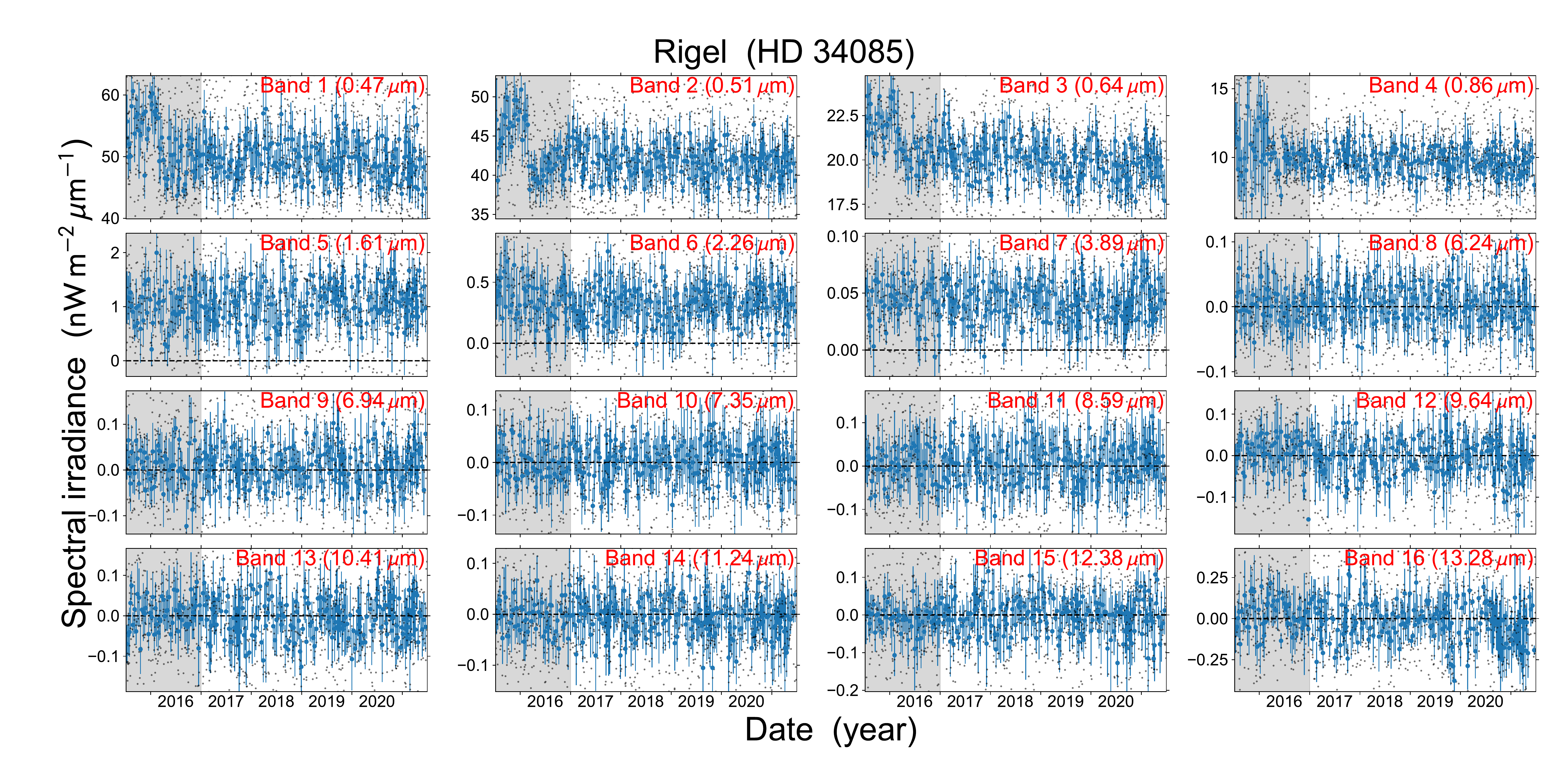}
\includegraphics[width=2\columnwidth]{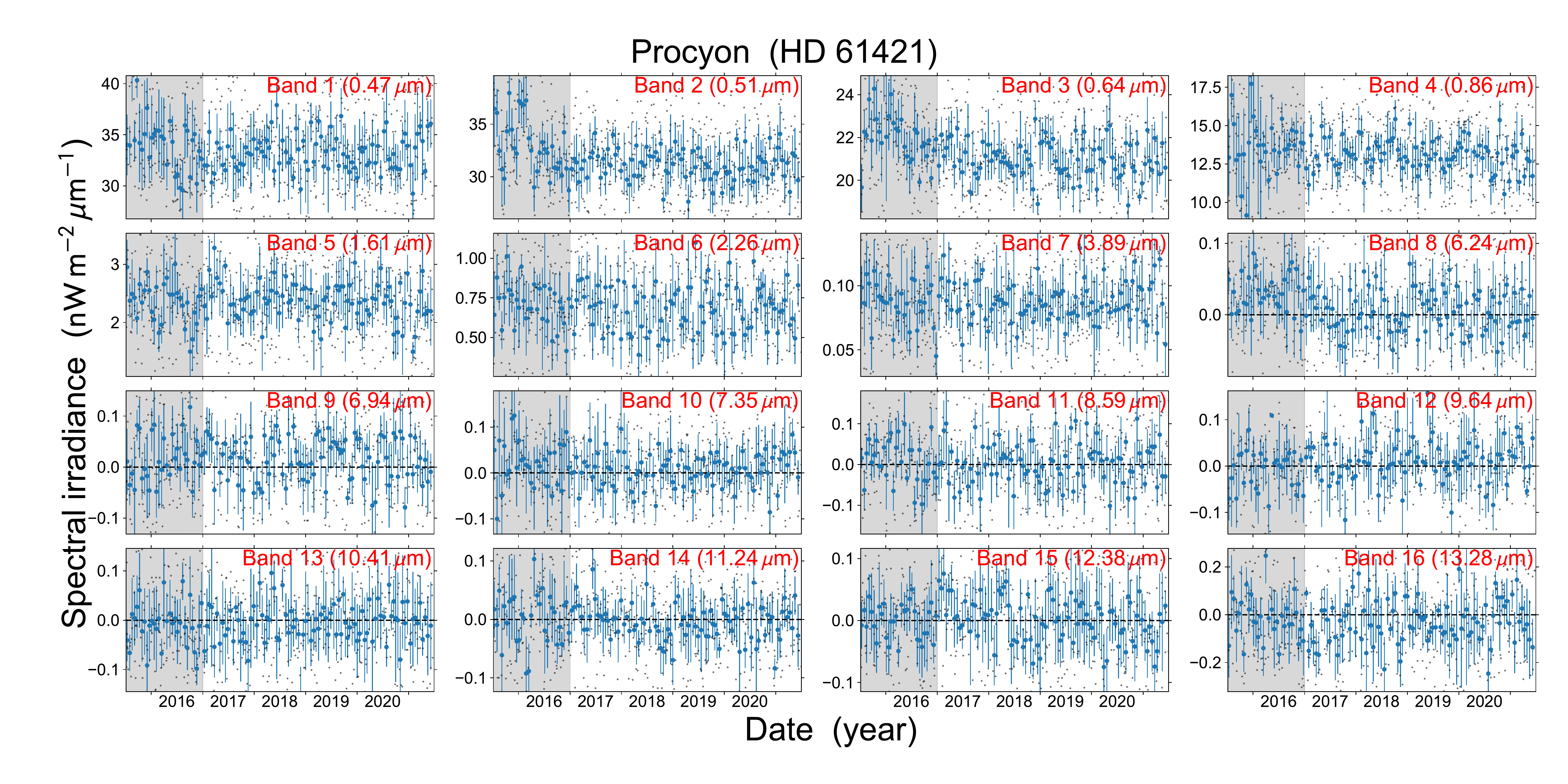}
\caption{\textbf{Light curves of Rigel and Procyon in all bands. } Same as Extended Data \autoref{fig:allbands_flux:Betelgeuse}\textbf{a}, but for the non-variable stars Rigel and Procyon. }
\label{fig:allbands_flux:Rigel-Procyon}
\end{figure*}

\begin{figure*}[t!]
\includegraphics[width=2\columnwidth]{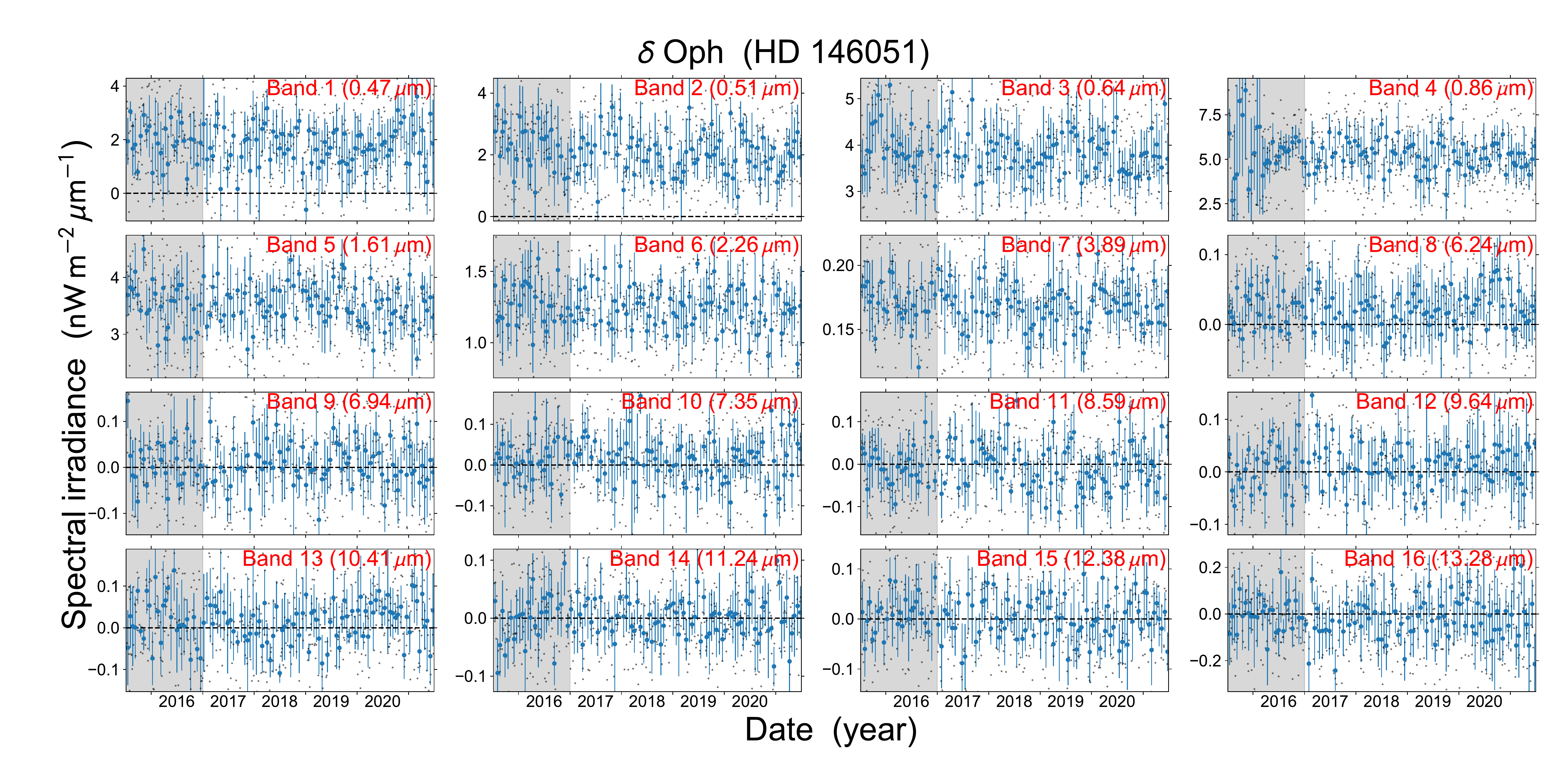}
\includegraphics[width=2\columnwidth]{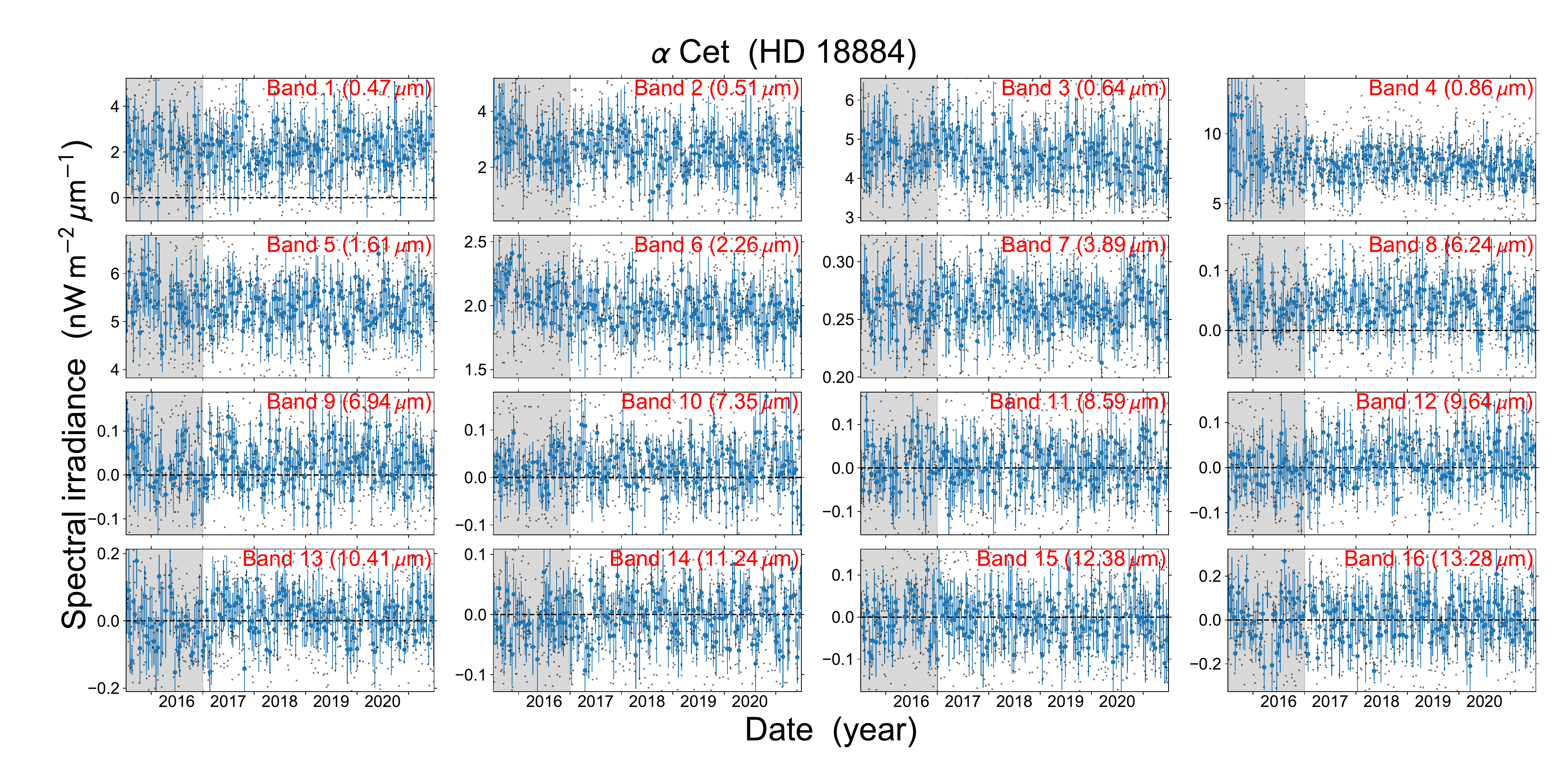}
\caption{\textbf{Light curves of $\delta $~Oph and $\alpha $~Cet in all bands. } Same as Extended Data \autoref{fig:allbands_flux:Betelgeuse}\textbf{a}, but for the red giant stars $\delta $~Oph and $\alpha $~Cet. }
\label{fig:allbands_flux:delOph-alfCet}
\end{figure*}

\begin{figure*}[t!]
\includegraphics[width=2\columnwidth ]{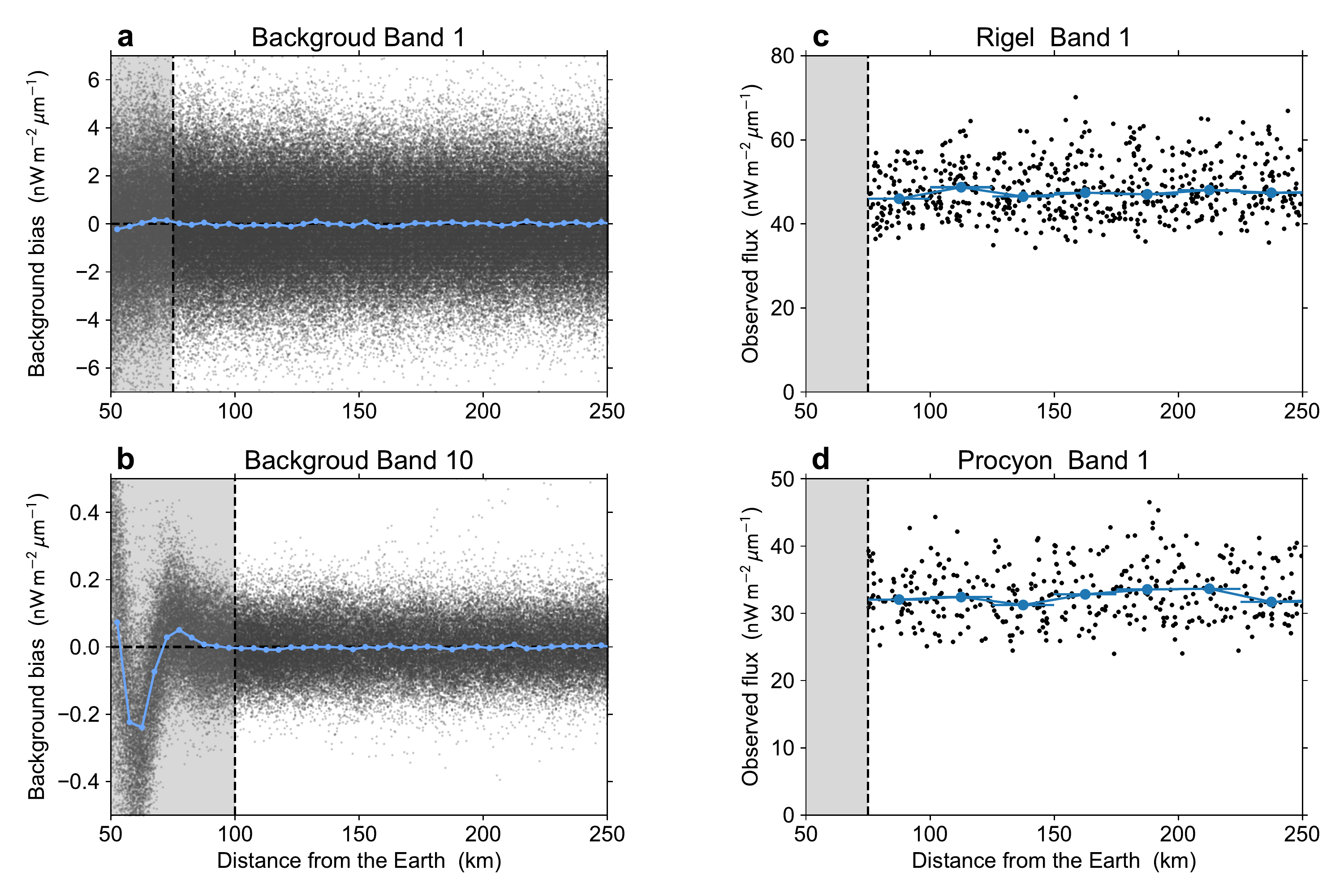}
\caption{\textbf{The effect of the Earth's atmosphere on the photometry. } Photometric measurements after the sky-removal procedures for the sky (\textbf{a} for Band~1 and \textbf{b} for Band~10) and for the non-variable stars in Band~1 (\textbf{c} for Rigel and \textbf{d} for Procyon) with respect to their distance from the edge of the Earth's disk. The scattered points represent individual measurements, and the solid lines denote the mean values in the bins. The standard errors of the mean values are shown with error bars (\textbf{c}, \textbf{d}), but they are too small to be visible. The vertical broken lines indicate the distance thresholds for each band. }
\label{fig:ray_height}
\end{figure*}

\begin{figure*}[t!]
\includegraphics[width=2\columnwidth ]{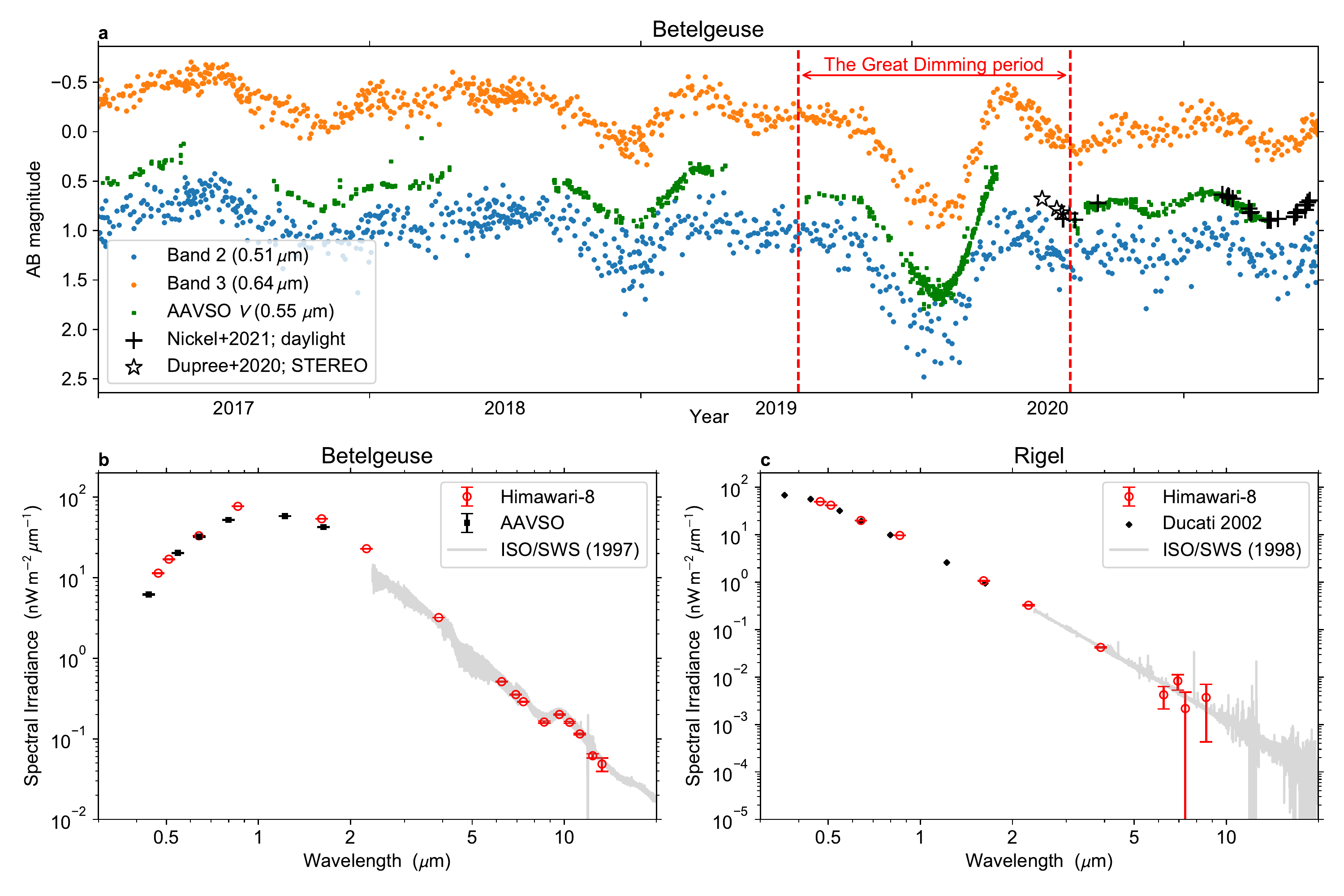}
\caption{\textbf{Photometric data of Betelgeuse and Rigel. } \textbf{a}, Comparison of the optical light curves of Betelgeuse. Blue and orange dots represent the optical light curves measured with Bands~2 and 3 of the Himawari-8 satellite, respectively. For comparison, green dots, black plus signs, and black stars represents the \textit{V}-band light curves taken from the literature; the AAVSO International Database, daylight observation\protect\citesi{Nickel2021_si}, and STEREO Mission data\protect\citesi{Dupree2020b_si}. \textbf{b} and \textbf{c}, SEDs of Betelgeuse and Rigel, respectively. The red circles represent our measurements with the Himawari-8 satellite after averaging all the observations from January 2017 to June 2021. Photometry from the literature is shown for comparison: the mean spectral irradiances between January 2017 and June 2021 taken from the AAVSO International Database (black squares in the left panel), those taken from the VisieR database\protect\citesi{Ducati2002_si} (black diamonds in the right panel), and infrared spectra observed with the ISO satellite\protect\citesi{Sloan2003_si} (gray lines). We ignored the photometric data for Betelgeuse during the Great Dimming period, August 2019--July 2020 indicated by vertical red dashed lines in \textbf{a}, in the averaging process. We calculated the spectral irradiances for these photometric data from the literature by converting from Vega to AB magnitudes using the zero points given in Table~1 in ref.\protect\citesi{Blanton2007_si}. We retrieved the AAVSO's photometric data on 16 February 2022. }
\label{fig:SED:Betelgeuse-Rigel}
\end{figure*}

\end{supplementaryinformation}

\end{document}